\documentclass[aps, physrev, reprint, superscriptaddress]{revtex4-1}

\usepackage{hyperref}
\usepackage{graphicx}
\usepackage[utf8]{inputenc}
\usepackage{amsmath, amssymb}

\date{\today}
\begin{document}

\title{Two-dimensional Mixture of Dipolar Fermions: Equation of State and Magnetic Phases}

\author{Tommaso Comparin}
\email{tommaso.comparin@unitn.it}
\affiliation{INO-CNR BEC Center and Dipartimento di Fisica, Universit\`a di Trento, 38123 Trento, Italy}

\author{Ra\'ul Bomb\'in}
\affiliation{Departament de F\'{i}sica, Universitat Polit\`{e}cnica de Catalunya, Campus Nord B4-B5, E-08034, Barcelona, Spain}
\affiliation{INO-CNR BEC Center and Dipartimento di Fisica, Universit\`a di Trento, 38123 Trento, Italy}

\author{Markus Holzmann}
\affiliation{Univ. Grenoble Alpes, CNRS, LPMMC, 3800 Grenoble, France}
\affiliation{Institut Laue-Langevin, BP 156, F-38042 Grenoble Cedex 9, France}

\author{Ferran Mazzanti}
\affiliation{Departament de F\'{i}sica, Universitat Polit\`{e}cnica de Catalunya, Campus Nord B4-B5, E-08034, Barcelona, Spain}

\author{Jordi Boronat}
\affiliation{Departament de F\'{i}sica, Universitat Polit\`{e}cnica de Catalunya, Campus Nord B4-B5, E-08034, Barcelona, Spain}

\author{Stefano Giorgini}
\affiliation{INO-CNR BEC Center and Dipartimento di Fisica, Universit\`a di Trento, 38123 Trento, Italy}

\begin{abstract}
We study a two-component mixture of fermionic dipoles in two dimensions at
zero temperature, interacting via a purely repulsive $1/r^3$ potential. This
model can be realized with ultracold atoms or molecules, when their dipole
moments are aligned in the confinement direction orthogonal to the plane.
We characterize the unpolarized mixture by means of the Diffusion Monte Carlo
technique. Computing the equation of state, we identify the regime of
validity for a mean-field theory based on a low-density expansion and
compare our results with the hard-disk model of repulsive fermions.
At high density, we address the possibility of itinerant ferromagnetism, namely
whether the ground state can be fully polarized in the fluid phase. Within the
fixed-node approximation,
we show that the accuracy of Jastrow-Slater trial wave functions, even with the typical
two-body backflow correction, is not sufficient to resolve the relevant energy
differences. By making use of the iterative-backflow improved trial
wave functions, we observe no signature of a fully-polarized ground state up to
the freezing density.
\end{abstract}

\maketitle

\section{Introduction}

The first experimental realizations of quantum-degenerate gases with ultracold
atoms made use of alkali species, for which interatomic interactions can be
treated as short-range potentials \cite{Dalfovo1999RMP, Giorgini2008RMP}. Other
atomic species possess a magnetic dipole moment which modifies the interactions
between particles \cite{Baranov2008PR, Lahaye2009RPP}, inducing a slow
large-distance decay and an anisotropic dependence on the atomic positions.
Alternative realizations of cold gases in the presence of dipolar interactions
are based on heteronuclear molecules with an induced electric dipole moment
\cite{Bohn2017} or on Rydberg atoms \cite{Santos2000PRL}.

Experiments in the presence of dipole-dipole interactions have lead to the
observation of peculiar phenomena, such as the Fermi-surface deformation in
Erbium \cite{Aikawa2014}, the long-range character of a dipolar Bose-Hubbard
model \cite{Baier2016} and quantum droplets \cite{Chomaz2016PRX,
FerrierBarbut2016PRL, Schmitt2016}. More recently, high experimental control
was reached in the preparation and characterization of dipolar mixtures, using
either different hyperfine states of the same element \cite{Baier2018PRL} or
different atomic species \cite{Trautmann2018PRL}. This route may lead to the
discovery of interesting many-body phenomena, including pairing and phase
separation \cite{Partridge2006} or dipolar magnetism \cite{StamperKurn2013RMP}.

The realization of quasi-two-dimensional configurations is experimentally
achieved through a strong confinement along the transverse direction.
This geometry may increase the stability of the system, as demonstrated for
ultracold gases of heteronuclear molecules \cite{Ni2010, DeMiranda2011NP}. The
theoretical study of two-dimensional models already provides a wealth of
predictions for many-body states with properties due to the shape of
dipole-dipole interaction. Consequences of the anisotropy include the stripe
phase of tilted dipolar bosons \cite{Macia2014PRA_1, Bombin2017PRL}, and the
stabilization of a fermionic superfluid with $p$-wave pairing
\cite{Bruun2008PRL, Cooper2009PRL}. Moreover, the slow decay of the repulsion at
large interatomic distances may lead to the formation of a solid phase
\cite{Buchler2007PRL, Astrakharchik2007PRL, Matveeva2012PRL}.

In this work, we compute ground-state properties of a binary mixture of fermions
with dipolar interactions, in a two-dimensional geometry. We focus on the case
of dipoles aligned along the transverse direction, which are characterized by an
isotropic $1 / r^3$ interparticle repulsion. While this system can be described
through mean-field theory in the low-density regime, quantum Monte Carlo has
been the only rigorous technique available to study fermionic dipoles at high
density. It was applied for instance to the fully-polarized state, namely a
system with a single species \cite{Matveeva2012PRL}. For such systems, a freezing
phase transition between a disordered and a solid phase takes place upon
increasing density.

We present the characterization of the unpolarized gas, both at low
densities (closer to the currently accessible experimental regimes) and at high
densities (relevant for fundamental questions about magnetic phases of
the mixture). Using the diffusion Monte Carlo (DMC) method \cite{Hammond1994,
Foulkes2001RMP, Kolorenc2011RPP}, we obtain accurate results for the equation of
state and pair distribution functions, which constitute a useful reference for
future experiments in quasi-two-dimensional geometries.

As a second part of our work, we consider the high-density regime and address
the possible occurrence of itinerant ferromagnetism, namely the existence of a
polarized ground state of the fluid. This phenomenon was first predicted for the
three-dimensional homogeneous electron gas \cite{Bloch1929ZP}, and represents a
long-standing topic in quantum many-body physics. It was the subject of
extensive theoretical investigations for the electron gas \cite{Ceperley1978PRB,
Ceperley1980PRL, Alder1982IJQC, Tanatar1989PRB, Ortiz1999PRL, Zong2002PRE,
Drummond2009PRL}, for liquid $^3$He \cite{Manousakis1983PRB, Zong2003MP,
Nava2012PRB} and for fermionic atoms with short-range interactions
\cite{Pilati2010PRL, Chang2011PNAS, AriasDeSaavedra2012PRA, Pilati2014PRL,
He2016PRA, Vermeyen2018PRA, Zintchenko2016EPJB}. The case of
short-range-interacting atoms is the closest to the textbook model of magnetism
introduced by Stoner \cite{Stoner1938PRSL}. Realized with a mixture of
ultracold $^6$Li atoms \cite{Valtolina2017NP}, it has been the only case where
an experimental signature of itinerant ferromagnetism was observed, despite the
difficulties related to the instability towards formation of two-body bound
states \cite{Jo2009, Pekker2011PRL, Sanner2012PRL, Amico2018PRL}.

The study of itinerant ferromagnetism has historically represented a challenge
and a testbed for many-body theories. Progress on this subject was especially
connected to technical advances in the field of DMC simulations, like the use of
backflow correlations and of twist-averaged boundary conditions -- see for
instance Ref.~\cite{Zong2002PRE}. In this work, we employ the DMC method for
the case of a two-dimensional dipolar gas, and we show that the level of accuracy
obtained with the commonly used Jastrow-Slater and backflow-corrected trial wave
functions is not sufficient to determine whether the ground state becomes polarized.
To go beyond this limitation, we make use of the recently-developed
iterative-backflow trial wave functions \cite{Taddei2015PRB, Ruggeri2018PRL},
finding no signature of a polarized ground state.

This article is structured as follows: In Section~\ref{sect:modelmethods} we
describe the model for two-dimensional dipoles and the Diffusion Monte Carlo
technique; Section~\ref{sect:P0} concerns the characterization of the
unpolarized mixture; and the issue of itinerant ferromagnetism at high density
is treated in Section~\ref{sect:ferromagnetism}.

\section{Model and methods}
\label{sect:modelmethods}

\subsection{Model for two-dimensional dipoles}

When dipoles are all aligned in the direction orthogonal to the plane of motion,
the dipole-dipole interaction is isotropic and decays as the inverse cube
distance between any two particles. We consider two fermionic species, labeled
as $\lbrace \uparrow, \downarrow \rbrace$ in analogy with the spin-$1/2$ case.
For $N = N_\uparrow + N_\downarrow$ particles in a square box of area $L^2$
(with periodic boundary conditions), the single-species density is $n_\sigma =
N_\sigma / L^2$ for $\sigma \in \left \lbrace \uparrow, \downarrow \right
\rbrace$, and the total density is $n = n_\uparrow + n_\downarrow = N / L^2$.
The $N$-particle Hamiltonian reads
\begin{equation}
H = -\frac{\hbar^2}{2m} \sum_{i=1}^N \nabla_i^2 + D \sum_{i<j} \frac{1}{r_{ij}^3},
\label{eq:Hamiltonian}
\end{equation}
where $r_{ij} \equiv |\mathbf{x}_i - \mathbf{x}_j|$ is the distance between
particles $i$ and $j$, and $D$ is proportional to the square of the dipole
moment \cite{Pitaevskii2016}. The characteristic length and energy scales are
given by $r_0 = m D / \hbar^2$ and $\varepsilon_0 = \hbar^2 / (m r_0^2)$. The
dimensionless density $n r_0^2$ encodes the strength of interactions.

The interparticle potential is species-independent. Since it extends beyond the
size of the simulation box, we apply a cut-off at distance $r_{ij} =
R_\mathrm{cut}$. The remaining contribution to the interaction energy is taken
into account by adding the \emph{tail correction}, $\Delta E_\mathrm{tail} =
\hbar^2 \pi
n r_0 / (m R_\mathrm{cut})$, which is computed assuming that the pair distribution
function is equal to one for distances $r > R_\mathrm{cut}$. For a large-enough
cutoff distance, this procedure yields results which do not depend on
$R_\mathrm{cut}$. When $R_\mathrm{cut} > L / 2$, the sum over particle pairs in
Eq.~\eqref{eq:Hamiltonian} is generalized to include particle images in
neighboring periodic boxes (\emph{cf.} for instance Refs.~\cite{Drummond2011PRB}
and \cite{Matveeva2012PRL}). To reduce the system-size dependence of the
energy, we also include the finite-size correction for the kinetic energy
of an ideal Fermi gas \cite{Lin2001PRE, Pilati2010PRL}. This is
especially relevant at low density, where the many-body state approaches the
ideal Fermi gas.

The population imbalance between the two species is kept fixed, and it is
encoded in the polarization,
\begin{equation}
P = \frac{N_\uparrow - N_\downarrow}{N_\uparrow + N_\downarrow},
\end{equation}
which we study for the $P = 1$ (fully-polarized) and $P = 0$ (unpolarized)
states.

\subsection{Diffusion Monte Carlo method}

We employ the Diffusion Monte Carlo (DMC) technique for finding the ground state
of the Hamiltonian in Eq.~\eqref{eq:Hamiltonian}. This is a stochastic method
that performs imaginary-time projection, which has been successfully applied to
many condensed-matter systems since it was developed \cite{Hammond1994,
Foulkes2001RMP, Kolorenc2011RPP}.

The systematic errors of DMC (due to the imaginary-time discretization and to
the use of a finite population of walkers) are kept under control, so that for
bosonic systems one deals with an exact determination of ground-state
properties within statistical noise. The use of a trial wave function $\Psi_T$
for importance sampling improves the efficiency of the estimation by reducing
the variance, without introducing an additional bias.

For fermionic systems, however, the presence of nodes in the wave function leads
to a sign problem which makes sampling intractable for large system size. This
is artificially cured by the fixed-node prescription \cite{Ceperley1980PRL,
Anderson1995IRPC}: After choosing the nodal structure of a trial many-body wave
function, the DMC technique is used to find the ground state under this
constraint, thus becoming a variational method. The fixed-node scheme
introduces an unknown systematic bias. This vanishes when the exact nodal
structure is used, which is in general not possible for two- or
three-dimensional systems.

We consider trial wave functions which are the product of two factors: A
Jastrow term $\Psi_S$, symmetric under the exchange of any two
particles, and a term $\Psi_A$ which is antisymmetric for the exchange of
particles of the same species, and which defines the nodal structure. The
simplest choices are the Jastrow-Slater (JS) and backflow (BF) wave functions.
In both cases the Jastrow term is a product of two-body correlation functions
(which we take as equal for all particle pairs, independently of their spin),
\begin{equation}
\Psi_S(R) = \prod_{i=1}^N \prod_{j<i} f_J(r_{ij}),
\end{equation}
where $R = \left\lbrace \mathbf{x}_1, \dots, \mathbf{x}_N \right\rbrace$ is the
collection of all particle coordinates. We adopt the choice for $f_J$ which
follows from combining the two-body solution for the $1/r^3$
interaction potential at short distances with the phononic long-range behavior
\cite{Astrakharchik2007PRL, Macia2011PRA, Matveeva2012PRL} for $r \le L / 2$,
namely
\begin{equation}
f_J(r) =
\begin{cases}
C_0 K_0\left( 2\sqrt{\frac{r_0}{r}} \right) & r< R_\mathrm{match},
\\[2mm]
C_1 \exp\left[ -\frac{C_2}{L - r} - \frac{C_2}{r} \right] & r>R_\mathrm{match},
\end{cases}
\end{equation}
where $K_0$ is the modified Bessel function. The parameters $C_0$, $C_1$, and
$C_2$ are fixed by imposing that both $f_J(r)$ and $f_J'(r)$ are continuous at
$r = R_\mathrm{match}$ and that $f_J(L/2)=1$. The distance $r =
R_\mathrm{match}$ where the two regimes have to match is a variational
parameter, which we optimize by minimizing the variational energy. Notice also
that $f_J'(L/2)=0$, in compliance with periodic boundary conditions.

For the antisymmetric part, we consider the product of two Slater determinants,
one for each species:
\begin{equation}
\Psi_A(R) =
\det M_\uparrow
\left( \mathbf{x}_1, \dots, \mathbf{x}_{N_\uparrow} \right)
\times
\det M_\downarrow
\left( \mathbf{x}_{N_\uparrow+1}, \dots, \mathbf{x}_{N} \right).
\label{eq:Psi_A}
\end{equation}
For the single-particle orbitals forming the matrices $M_\sigma$ the simplest
choice is to use plane waves, $\exp\left( i\,\mathbf{k}_m \cdot \mathbf{x}_n
\right)$, and to fill the momentum states $\mathbf{k}_m$ associated to the
periodic box according to their increasing energy. This choice leads to the JS
trial wave function, $\Psi_\mathrm{JS}$, which has the same nodes of the ideal
Fermi gas. To improve the nodal structure, a backflow correction can be
included, which consists in replacing the argument $\mathbf{x}_i$ of a
plane-wave single-particle orbital with a function of all the particle coordinates:
\begin{equation}
\mathbf{x}_i
\rightarrow
\mathbf{q}_i \equiv \mathbf{x}_i +
\sum_{j\neq i} \left(\mathbf{x}_i - \mathbf{x}_j \right) f_\mathrm{BF}(r_{ij}).
\label{eq:BF}
\end{equation}
In $\Psi_\mathrm{BF}$, we use the same parametrization for $f_\mathrm{BF}$ as in
Ref.~\cite{Grau2002PRL}, namely a Gaussian function $f_\mathrm{BF}(r) = C_3
\exp\lbrace-[ (r - C_4)/C_5]^2 \rbrace$. The free parameters $C_3$, $C_4$ and
$C_5$ are set by optimizing the variational energy of the trial wave function,
while controlling the width of the Gaussian so that $f_\mathrm{BF}(L/2)$ remains
small enough to avoid boundary effects.

While being sufficiently accurate for many purposes, especially at low density,
the JS and BF trial wave functions are not sufficient for the study of
itinerant ferromagnetism, which we present in Section~\ref{sect:ferromagnetism}.
The possible ways to further improve the nodes of $\Psi_T$ include an explicit
three-body backflow correction \cite{Casulleras2000PRL, Holzmann2006PRB}, and
the recent proposal of using iterative-backflow wave functions
\cite{Taddei2015PRB, Ruggeri2018PRL}. In the latter case, one constructs a
sequence $\alpha=0,1, 2, ...$ of backflow transformations
$f_\mathrm{BF}^{(\alpha)}$ [\emph{cf}. Eq.~\eqref{eq:BF}], starting from the
physical coordinates $\mathbf{q}_i^{(0)} \equiv \mathbf{x}_i$ and iteratively
obtaining $\lbrace \mathbf{q}^{(\alpha)}_i \rbrace_{i=1}^N$ from $\lbrace
\mathbf{q}_i^{(\alpha - 1)} \rbrace_{i=1}^N$. We choose a suitable parametrization for
the functions $f_\mathrm{BF}^{(\alpha)}$, and fix their free parameters by
optimizing the variational energy $E_T$ -- see
Appendix~\ref{app:iterative_backflow} for full details.

The most directly accessible observable within DMC is the total energy, which
constitutes an upper bound to the true ground-state energy, within the
fixed-node scheme. We compute pair distribution functions through the
forward-walking technique \cite{Casulleras1995PRB} to obtain pure Monte Carlo
estimators (that is, free of systematic errors, apart from the fixed-node
approximation).

\section{Unpolarized mixture: Equation of state and pair distribution functions}
\label{sect:P0}

In this section we characterize the $P = 0$ state of dipolar fermions in two
dimensions, by computing its energy and pair distribution functions at different
densities.

\subsection{Equation of state: Analytical approximations}
\label{sect:P0:eos}

In the low-density regime, $n r_0^2 \ll 1$, approximate theories are
available for the ground-state energy of the Hamiltonian in
Eq.~\eqref{eq:Hamiltonian}. The starting point is the ideal Fermi gas (IFG),
without dipolar repulsion ($D = 0$), which has an energy per particle
$E_\mathrm{IFG}$ given by
\begin{equation}
\frac{E_\mathrm{IFG}}{\hbar^2/m}
=
\frac{1 + P}{2} \pi n_\uparrow + \frac{1 - P}{2} \pi n_\downarrow
=
\frac{\pi n}{2} \left( 1 + P^2 \right),
\label{eq:eos:ifg}
\end{equation}
in the thermodynamic limit. The contribution of dipolar interactions can be
included through approximate schemes, valid at low density. For a
fully-polarized state ($P = 1$), Hartree-Fock (HF) theory yields an expression
for the first correction $E_\mathrm{HF}$ to the IFG energy,
\begin{equation}
\frac{E_\mathrm{HF}}{E_\mathrm{IFG}} =
\frac{256}{45\sqrt{\pi}} \sqrt{n r_0^2},
\label{eq:eos:p1:hf}
\end{equation}
while higher-order terms may be obtained through many-body perturbation theory
\cite{Lu2012PRA}.

The Fourier transform of $1/r^3$ in two dimensions is not well-defined, unless a
regularization scheme is used -- see for instance Ref.~\cite{Lange2016PRA}. In
the derivation of Eq.~\eqref{eq:eos:p1:hf}, a cancellation between the direct
and exchange contributions to the Hartree-Fock energy solves this issue, but the
same cancellation does not take place when two different fermionic species are
considered. Therefore, for $P < 1$, Hartree-Fock theory cannot be directly applied to
dipolar gases described by the model in Eq.~\eqref{eq:Hamiltonian}.

To treat a two-component mixture we use a different approach, where we replace
the dipolar repulsion between particles of different species with a zero-range
interaction. This replacement is only valid in the low-density limit, where
properties of the gas should depend only on the two-dimensional scattering
length $a_s$, and not on the microscopic details of the interactions. For the
$1/r^3$ repulsion, the two-dimensional scattering length reads $a_s = r_0
e^{2\gamma}$ \cite{Ticknor2009PRA, Macia2011PRA}, with $\gamma \simeq 0.5772$ the
Euler constant.
Thus the approximate expression for the energy per particle of an unpolarized mixture ($P
= 0$) is
\begin{equation}
E = E_\mathrm{IFG} + E_\mathrm{MF},
\label{eq:eos:p0}
\end{equation}
as in the case of zero-range interactions.
The mean-field (MF) interaction energy reads \cite{Schick1971PRA}
\begin{equation}
\frac{E_\mathrm{MF}}{\hbar^2/m} = \frac{\pi n}{\left| \log \left(c_0 n a_s^2 \right) \right|},
\label{eq:eos:p0:mf}
\end{equation}
and it depends on a free parameter $c_0$.
In Eq.~\eqref{eq:eos:p0}, $E_\mathrm{MF}$ only accounts for interactions between
$\uparrow$ and $\downarrow$ particles. Notice that the Hartree-Fock
contributions $E_\mathrm{HF}$ for same-species repulsion is proportional to
$n^{3/2}$ [see Eq.~\eqref{eq:eos:p1:hf}], and in the limit $n r_0^2 \to 0$ it
yields a subleading correction with respect to $E_\mathrm{MF}$. Including
$E_\mathrm{HF}$ in Eq.~\eqref{eq:eos:p0} would constitute an uncontrolled
approximation, since it is unknown how $E_\mathrm{HF}$ would combine with the
beyond-mean-field correction for the opposite-spin interaction energy
\cite{Astrakharchik2009PRA}.

The dependence of $E_\mathrm{MF}$ on the free parameter $c_0$ is a peculiarity
of the two-dimensional case, where the mean-field coupling constant is only
known up to logarithmic accuracy \cite{Pitaevskii2016}. Following
Ref.~\cite{Bertaina2013EPJ}, we set $c_0 = (\pi / 2) \exp(2\gamma) \simeq
4.9829$, which corresponds to setting an energy scale equal to twice the Fermi
energy of the unpolarized case. For the beyond-mean-field contribution, several
expressions have been proposed and tested for the bosonic case
\cite{Pilati2005PRA, Astrakharchik2007PRA, Astrakharchik2009PRA}.

\subsection{Equation of state: DMC numerical results}

For the polarized state, the Hartree-Fock prediction of Eq.~\eqref{eq:eos:p1:hf} was
already compared with the DMC equation of state in Ref.~\cite{Matveeva2012PRL}, finding
that it is a valid approximation to the ground-state energy at densities up to
$n r_0^2 \approx 10^{-2}$. Here we compute the DMC energy per particle for an
unpolarized mixture ($P = 0$) and compare it with the expansion in
Eq.~\eqref{eq:eos:p0}.

\begin{figure}[htb]
\centering
\includegraphics[width=\linewidth]{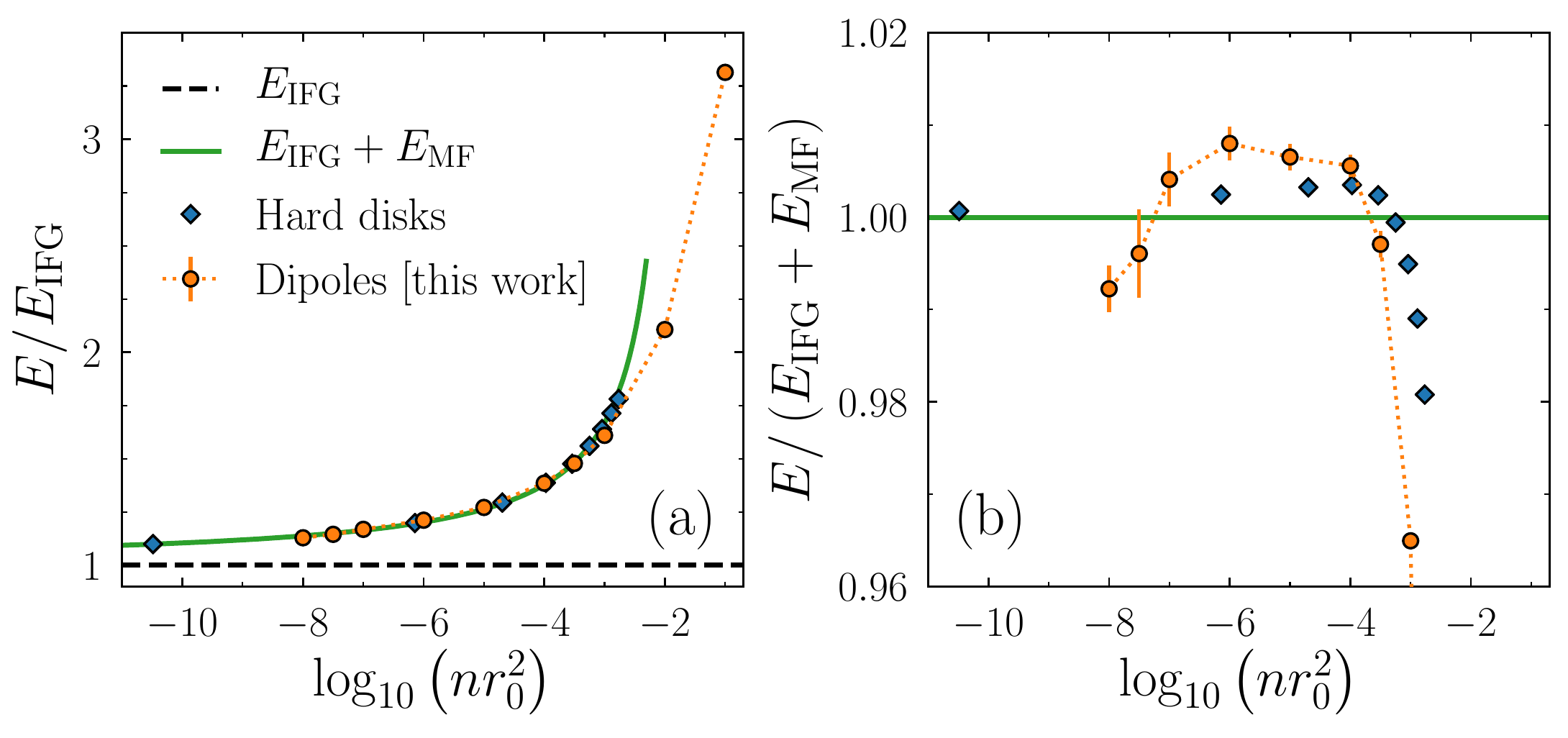}
\caption{(a) Equation of state for the unpolarized system ($P = 0$) in units of
the ideal-Fermi-gas energy $E_\mathrm{IFG}$. DMC data for the dipolar gas
(orange circles -- orange dotted lines are a guide to the eye) are compared to
the IFG and MF curves (black dashed line and green solid line) and to the DMC
energy for hard disks (blue diamonds, from Ref.~\cite{Bertaina2013EPJ}, with
disk diameter $a_s = e^{2\gamma} r_0$). (b) Same as in panel (a), in units of
the MF equation of state. In panel (a), statistical uncertainties on the DMC
energies (orange circles) are smaller than the symbol size.
}
\label{fig:eos}
\end{figure}

The DMC equation of state (obtained using $\Psi_\mathrm{JS}$ as trial wave
function) is shown in Fig.~\ref{fig:eos}(a). The ground-state energy of the
dipolar gas is systematically larger than the ideal-Fermi-gas value
$E_\mathrm{IFG}$, consistently with the fully-repulsive nature of interactions.
At density as low as $n r_0^2 = 10^{-8}$, the relative energy deviation from
$E_\mathrm{IFG}$
is still above 10\%.
The mean-field equation of state, in contrast, reproduces the DMC
energy with a higher accuracy -- see Fig.~\ref{fig:eos}(b).
Clear deviations from
the MF curve appear for densities $n r_0^2 \gtrsim 10^{-3}$.

Also the case of a single-species bosonic gas of dipoles in two dimensions was
studied through DMC, to compute the equation of state and compare it with a
mean-field approximation \cite{Astrakharchik2009PRA}. High-precision energies
were obtained for densities down to $n r_0^2 = 10^{-100}$, where the
beyond-mean-field energy correction remains as large as 1\%. An analogous
precision, which is necessary to properly describe the subtle beyond-mean-field
contribution, is beyond the scope of the current work.

Apart from the benchmark of approximate theories, it is instructive to compare
the low-density equation of state for dipoles with the one for a different model
of repulsive fermions, namely hard disks. In the case of bosons, a universal
regime exists for densities below $n r_0^2 \approx 10^{-7}$, where the energy of
these two models only depends on the gas parameter $n a_s^2$
\cite{Pilati2005PRA, Astrakharchik2009PRA}. To establish such comparison in the
fermionic case, we consider hard disks with diameter and scattering length both
equal to the scattering length of dipoles, $a_s = r_0 e^{2\gamma}$. The DMC
equation of state for hard disks from Ref.~\cite{Bertaina2013EPJ} is in
reasonable agreement with the one for dipoles [see Fig.~\ref{fig:eos}(a)]. This
behavior hints at the presence of a universal low-density regime, where the
energy only depends on the gas parameter $n a_s^2$. Nevertheless, we observe a
small deviation between the energies of dipoles and hard disks, highlighted in
Fig.~\ref{fig:eos}(b).
The residual difference suggests that the universal
region would start at even lower density, when compared to the hard-disk
case.
This is because the dipolar potential, in spite of its
formal short-ranged nature, has a slow power-law decay at large distance
[see also the discussion about
pair distribution functions in Section~\ref{sect:P0:gr}]. The effect of higher
partial waves may also contribute to the energy deviations.

Notice that backflow corrections are negligible for computing the equation of
state at low density, as in Fig.~\ref{fig:eos}. They become important at high
density, especially when one needs to evaluate the small energy difference
between the polarized and unpolarized states. This is the case for the study of
itinerant ferromagnetism, where the bias stemming from the fixed-node
prescription becomes crucial -- see Section~\ref{sect:ferromagnetism}.

\subsection{Pair distribution functions}
\label{sect:P0:gr}

\begin{figure}[hbt]
\centering
\includegraphics[width=\linewidth]{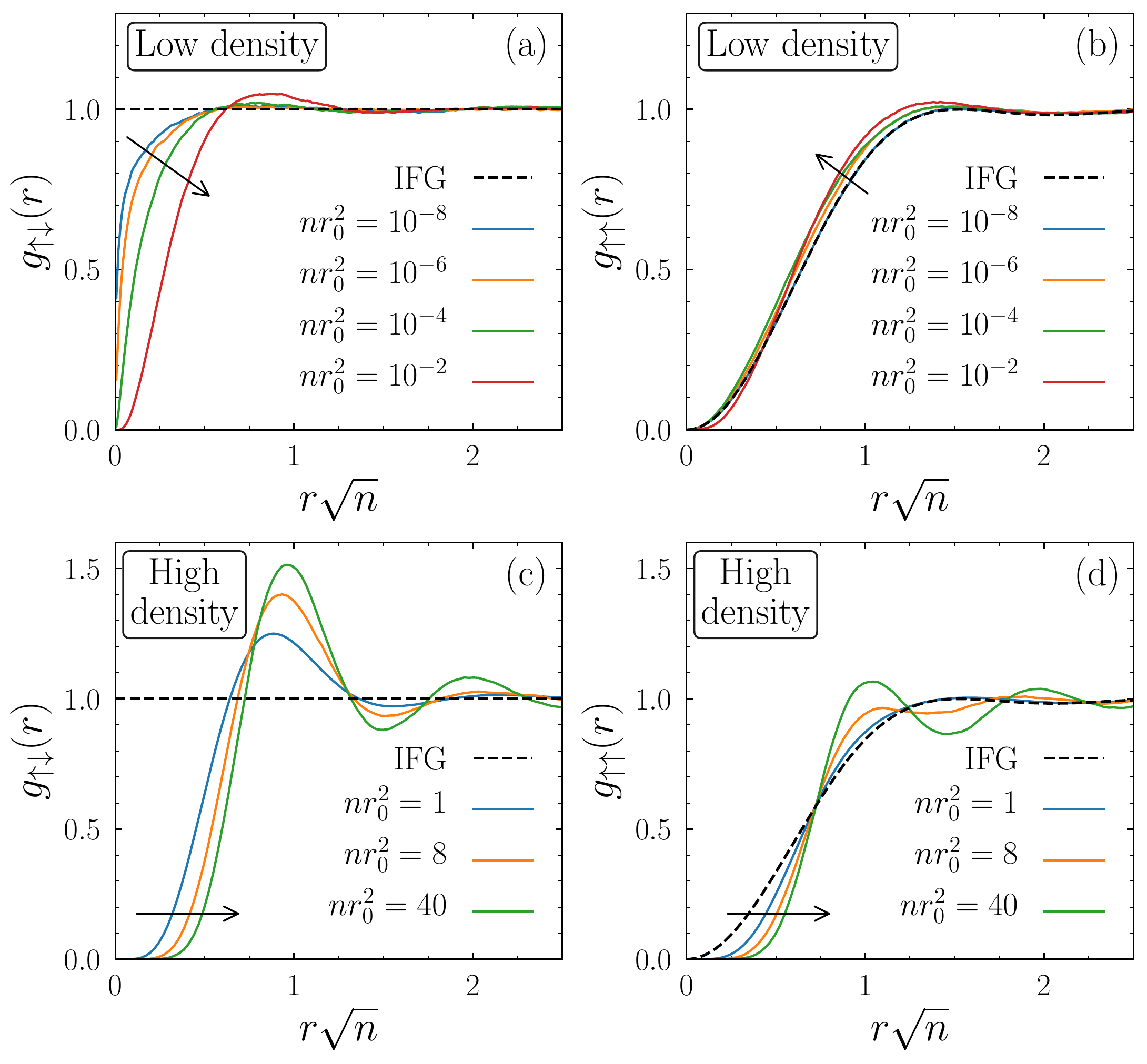}
\caption{Pair distribution functions for the $P=0$ state, computed through DMC
(solid lines), for different-species [panels (a) and (c)] and same-species pairs
[panels (b) and (d)]. The analytic IFG distribution
functions are also shown (dashed black lines). Arrows point towards increasing
$nr_0^2$. The particle numbers are $N = 74$ (for $n r_0^2 < 1$), $N = 98$ (for
$n r_0^2 = 1$) and $N = 122$ (for $n r_0^2 > 1$). The trial wave functions are
$\Psi_\mathrm{JS}$ for $n r_0^2 < 1$ and $\Psi_\mathrm{BF}$ for $n r_0^2 \geq
1$. Statistical error bars are of the order of the line width.}
\label{fig:gr}
\end{figure}

As part of the characterization of the balanced mixture of fermionic dipoles, we
compute the same-species and different-species pair distribution functions
[$g_{\uparrow\uparrow}(r)$ and $g_{\uparrow\downarrow}(r)$, respectively] for
both low and high densities -- see Fig.~\ref{fig:gr}. As a reference, we compare
our results to the equivalent curves for the IFG: $g_{\uparrow\downarrow}(r) =
1$ and $ g_{\uparrow\uparrow}(r) = 1 - \left(2 J_1(\tilde{r}) /
\tilde{r}\right)^2$, where $\tilde{r} \equiv (4 \pi n_\uparrow)^{1/2} r$ and
$J_1$ is the Bessel function of the first kind \cite{Giuliani2005}.

At low density, $g_{\uparrow\downarrow}(r)$ is similar to the IFG curve, apart
from a short-distance suppression due to dipolar repulsion -- \emph{cf.}
Fig.~\ref{fig:gr}(a). In the same density regime, $g_{\uparrow\uparrow}(r)$
displays a different behavior [see Fig.~\ref{fig:gr}(b)]: At density $n r_0 =
10^{-8}$ it is indistinguishable from the corresponding IFG curve, and for
increasing densities (up to $n r_0^2 \lesssim 10^{-4}$) it shifts towards
smaller values of $r \sqrt{n}$. The $g_{\uparrow\uparrow}(r)$ curve for $n
r_0^2 = 10^{-2}$, in contrast, intersect with the ones for lower densities
[\emph{cf}. red line in Fig.~\ref{fig:gr}(b)]. We interpret this change of
behavior as the departure from the weakly-interacting regime.

\begin{figure}[htb]
\centering
\includegraphics[width=0.9\linewidth]{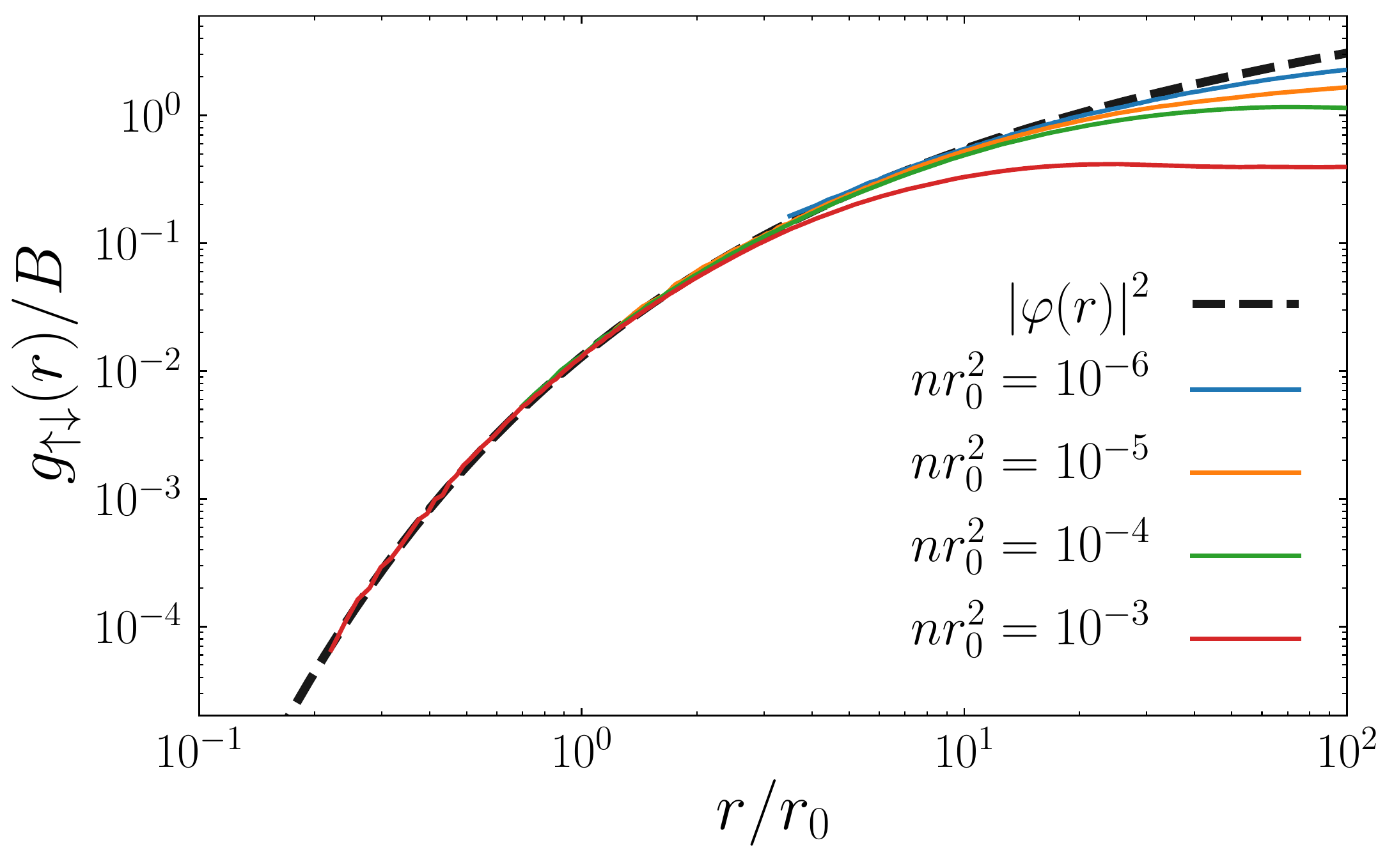}
\caption{Pair distribution function $g_{\uparrow\downarrow}(r)$ at small
distance, for different densities (solid lines from top to bottom are for
densities from $n r_0^2 = 10^{-6}$ to $n r_0^2 = 10^{-3}$). Each curve is
fitted to $B |\varphi(r)|^2$ in the small-$r$ regime, and plotted in units of
$B$ to highlight the collapse onto the squared two-body wave function
$|\varphi(r)|^2$ (black dashed line). Notice that the value of $B$ depends on
the density $n r_0^2$. Statistical error bars are of the order of the line
width.}
\label{fig:gr:smallr}
\end{figure}

At low-enough density $n r_0^2$ or at small-enough distance $r$, the
probability of having three particles at distance $r$ from each other is
suppressed, and $g_{\uparrow\downarrow}(r)$ only depends on two-body properties.
The solution of the Schr\"{o}dinger equation for two dipoles in vacuum
is $ \varphi(r) \propto K_0 ( 2 \sqrt{ r_0 / r } )$, with $K_0$ the modified
Bessel function \cite{Macia2011PRA}. Our DMC data confirm that the
$g_{\uparrow\downarrow}(r)$ curves for different densities collapse onto
$|\varphi(r)|^2$, at short distance, after the rescaling by a density-dependent
constant prefactor $B$ -- see Fig.~\ref{fig:gr:smallr}. Deviations from this
behavior appear at a distance which is proportional to $1/\sqrt{n}$, so that the
two-body window becomes smaller for higher densities.

Despite its slow decay at large distance, the $1 / r^3$ interaction is formally
short-ranged in two dimensions. The wave function $\varphi(r)$ for the two-body
problem in vacuum scales as $\log r $ at large $r$, as for other short-range
models. For such potentials there exists a universal
window where $g_{\uparrow\downarrow}(r) \propto \log^2 r$, for
distances much larger than the potential range and much smaller than the typical
interparticle distance. The constant prefactor of $\log^2 r$ is Tan's contact
parameter, and a set of universal expressions connects it to other
observables \cite{Werner2012PRA_1}. This behavior is expected also for
the dipolar gas, but the slow decay of $1/r^3$ is reflected in a slow convergence
of $\varphi(r)$ towards its asymptotic form.
The universal window is not present for the densities considered in this work,
and would only appear for higher values of $1/\sqrt{n}$ (that is, for lower
density). Thus it is not feasible to fit a $\log^2 r$ curve to the data in
Fig.~\ref{fig:gr:smallr}, at
odds with the case of short-range potentials like hard disks \cite{Bertaina2013EPJ}.
This non-universality
is consistent with the observed deviations of the equation of state for the
dipolar gas from the one for
hard disks (see Fig.~\ref{fig:eos}).
The universal regime would appear at even lower densities, which are unrealistic
for current experiments.

Also in the high-density regime [$n r_0^2 \geq 1$, see Fig.~\ref{fig:gr}(c) and
(d)], $g_{\uparrow\downarrow}(r)$ reflects the dipolar repulsion,
showing the oscillatory behavior that corresponds to the formation of shells of
neighbors around a specific particle. As the interaction strength increases, the
height of the first peak becomes larger. Particles of the same
species, on the contrary, are kept farther apart due to the Pauli exclusion
principle. A direct consequence is
that the first peak of $g_{\uparrow\uparrow}(r)$ is suppressed, as compared to the
one in $g_{\uparrow\downarrow}(r)$.
At $n
r_0^2 = 1$ the first peak is invisible, at $n r_0^2 = 8$ it is visible but lower
than the second one, and at $n r_0^2 = 40$ the two are comparable [see
Fig.~\ref{fig:gr}(d)].

The effect of using $\Psi_\mathrm{BF}$ rather than $\Psi_\mathrm{JS}$, when
computing pair distribution functions within DMC, is negligible for densities $n
r_0^2 \ll 1$ (at the scale considered). For the other cases reported in
Fig.~\ref{fig:gr}, with $nr_0^2 \geq 1$, we explicitly evaluated the backflow
correction to $g_{\sigma_1 \sigma_2}(r)$ through the comparison of results
obtained by using $\Psi_\mathrm{BF}$ or $\Psi_\mathrm{JS}$. At the largest
density ($n r_0^2 = 40$), the maximum absolute effect is a variation of $0.04$
close to the first peak, both for $g_{\uparrow\uparrow}(r)$ and
$g_{\uparrow\downarrow}(r)$. This corresponds to less than a 6\% relative change
in the pair distribution functions.

\section{Itinerant ferromagnetism}
\label{sect:ferromagnetism}

In Sections~\ref{sect:ferromagnetism:direct} and
\ref{sect:ferromagnetism:polaron}, we present two standard methods to look for a
ferromagnetic instability, both based on commonly used trial wave functions
($\Psi_\mathrm{JS}$ and $\Psi_\mathrm{BF}$): The direct comparison of the energy
for many-body states with $P=0$ and $P=1$, and the comparison of the polaron
energy with the chemical potential of the fully-polarized gas. The inconclusive numerical results
constitute a motivation for resorting to more accurate trial wave functions for
DMC, as reported in Section~\ref{sect:ferromagnetism:iterativeBF}.

\subsection{Direct energy comparison for $P = 0$ and $P = 1$}
\label{sect:ferromagnetism:direct}

A direct method to look for a ferromagnetic instability is to compute the
difference between the energy per particle of the unpolarized and polarized
states,
\begin{equation}
\Delta E \equiv E_{P=0} - E_{P=1}.
\label{eq:DeltaE}
\end{equation}
The two states are simulated by using trial wave functions with different nodal
structures. For $P = 0$ the antisymmetric part of $\Psi_T$ is the product of two
Slater determinants [one for each species, as in Eq.~\eqref{eq:Psi_A}], while a
single one is needed for $P = 1$.

\begin{figure}[htb]
\centering
\includegraphics[width=\linewidth]{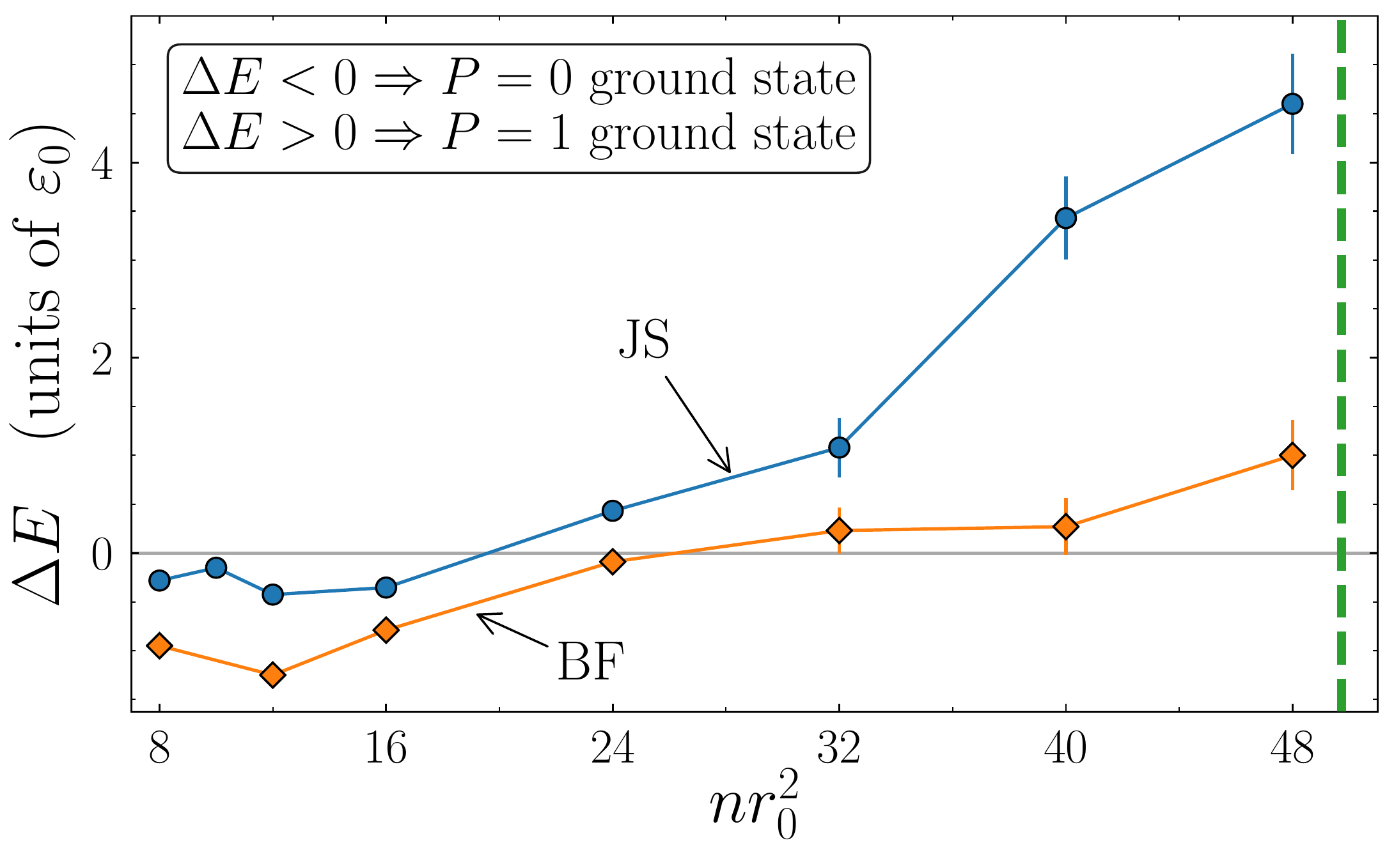}
\caption{Difference in the energy per particle between $P=0$ and $P=1$ states
[\emph{cf.} Eq.~\eqref{eq:DeltaE}], computed through DMC with the Jastrow-Slater
(JS) or Backflow (BF) trial wave function (see Table~\ref{tab:energies_JS_BF}).
Statistical error bars are shown, and lines are a guide to the eye. The vertical
dashed line marks the freezing density for the $P = 1$ state
\cite{Matveeva2012PRL}.}
\label{fig:DE_P0_P1}
\end{figure}

We compute $\Delta E$ twice, with the SJ and BF trial wave
functions. In both cases we find a critical density at which $\Delta E$ crosses
zero and becomes positive, which signals a region where the system features a
magnetic instability towards a polarized state -- see Fig.~\ref{fig:DE_P0_P1}.
This crossing point only provides an upper bound for the onset of
ferromagnetism, since in principle a partially-polarized state (with $0 < P <
1$) could be the ground state even at lower density.

The crossing between $P=0$ and $P=1$ takes place at density $n r_0^2$
approximately equal to 20 when using $\Psi_\mathrm{JS}$, and to 26 when using
$\Psi_\mathrm{BF}$ -- see Fig.~\ref{fig:DE_P0_P1} and
Table~\ref{tab:energies_JS_BF}. The shift of the crossing density is mainly due
to the large backflow correction to the $P=0$ energy [see fourth and seventh
columns in Table~\ref{tab:energies_JS_BF}], as in the case of the
three-dimensional electron gas \cite{Zong2002PRE}. It is known from
previous calculations on liquid $^3$He that spherically-symmetric backflow correlations have a small
effect in the polarized phase
\cite{Pandharipande1973PRC, Manousakis1983PRB}. For the
dipolar gas, the relative energy difference between $P = 1$ and $P = 0$ is
extremely small (below $5\cdot10^{-4}$, for calculations with $\Psi_\mathrm{BF}$
at density $n r_0^2$ between $24$ and $48$). These two elements (the clear shift
in the crossing density and the small energy differences) suggest that the
fixed-node bias is crucial and may lead to qualitatively different results when
using a different nodal structure, as we show in
Section~\ref{sect:ferromagnetism:iterativeBF}. This is especially important
because the relevant density window for a hypothetical polarized phase has an
upper bound at $n r_0^2 \approx 50$, the freezing density for the $P = 1$ state
\cite{Matveeva2012PRL}.

We stress that data in Table~\ref{tab:energies_JS_BF} and
Fig.~\ref{fig:DE_P0_P1} are obtained for systems of finite size ($N = 121,
122$), chosen as large as possible to minimize the size dependence of both the
kinetic and potential energies. A study of the scaling of $E$ with respect to
$N$ reveals that the residual finite-size dependence of the energies does not
modify the result about the existence of the $\Delta E = 0$ energy crossing.

\begin{table}[htb]
\begin{center}
\begin{tabular}{c | c c c | c c c}
& & $P = 1$ &
& & $P = 0$ \\
$n r_0^2$ &
$E_\mathrm{JS}$ & $E_\mathrm{BF}$ & Corr. &
$E_\mathrm{JS}$ & $E_\mathrm{BF}$ & Corr. \\
\hline
 8 & 168.66(1) & 168.63(1) & 0.03(1) & 168.38(2) & 167.68(4) & 0.70(4)\\
12 & 295.98(2) & 295.92(2) & 0.06(3) & 295.55(6) & 294.67(5) & 0.88(8)\\
16 & 442.22(2) & 442.04(2) & 0.18(3) & 441.87(7) & 441.25(6) & 0.6(1)\\
24 & 781.21(4) & 780.8(1) & 0.4(1) & 781.6(1) & 780.71(7) & 0.9(1)\\
32 & 1172.45(5) & 1171.9(1) & 0.6(1) & 1173.5(3) & 1172.1(2) & 1.4(4)\\
40 & 1608.0(1) & 1607.7(1) & 0.3(2) & 1611.4(4) & 1607.9(2) & 3.5(5)\\
48 & 2083.1(1) & 2083.0(2) & 0.1(2) & 2087.7(5) & 2084.0(3) & 3.7(6)\\
64 & 3137.7(1) & 3137.4(2) & 0.3(2) & 3145.4(8) & 3139.8(3) & 5.6(9)\\
\end{tabular}
\end{center}
\caption{DMC energy per particle (in units of $\varepsilon_0$) at different
densities.
Data for $P=1$ ($P=0$) are obtained with $N = 121$ ($N = 122$) particles.
Columns marked as ``Corr.'' indicate the energy correction $E_\mathrm{JS} -
E_\mathrm{BF}$.
Statistical errors are reported in parentheses.
\label{tab:energies_JS_BF}}
\end{table}

\subsection{Stability of $P = 1$ state}
\label{sect:ferromagnetism:polaron}

An alternative approach to look for itinerant ferromagnetism consists in
computing the energy cost or gain for a spin-flip excitation on top of a fully
polarized state. If the $P = 1$ state is the ground state, then any excitation
must lead to a state with higher energy. If the true ground state has
polarization $P < 1$, on the contrary, a single spin flip on top of a
fully-polarized state may decrease its energy, signalling the instability of the
$P = 1$ state.

Through the DMC technique, we have access to the energy per particle
$E_{(N_\uparrow, N_\downarrow)}$, for any choice of $(N_\uparrow,
N_\downarrow)$. For a given $N_\uparrow$, the polaron energy $\varepsilon_p$ and
the $P=1$ chemical potential $\mu$ read
\begin{align}
\varepsilon_p^{(N_\uparrow)} & \equiv
(N_\uparrow + 1) E_{(N_\uparrow, 1)} - N_\uparrow E_{(N_\uparrow,0)},\\
\mu^{(N_\uparrow)} & \equiv
(N_\uparrow + 1) E_{(N_\uparrow + 1, 0)} - N_\uparrow E_{(N_\uparrow, 0)},
\label{eq:epsP_and_mu}
\end{align}
where both quantities are defined at fixed surface $L^2$ (notice that this
leads to a density difference between systems of $N_\uparrow$ and $N_\uparrow+1$
particles, which vanishes as $1 / N_\uparrow$ in the thermodynamic limit).

In a large system, a single spin flip on top of a fully-polarized state induces a change
in the total energy which is
\begin{equation}
\Delta E_\mathrm{flip} = \varepsilon_p^{(\infty)} - \mu^{(\infty)}.
\end{equation}
If $\Delta E_\mathrm{flip} > 0$, then the $P = 1$ state is robust against a
single spin-flip excitation and might be the true ground state, while for a
negative $\Delta E_\mathrm{flip}$ such state is unstable.

\begin{figure}[htb]
\centering
\includegraphics[width=\linewidth]{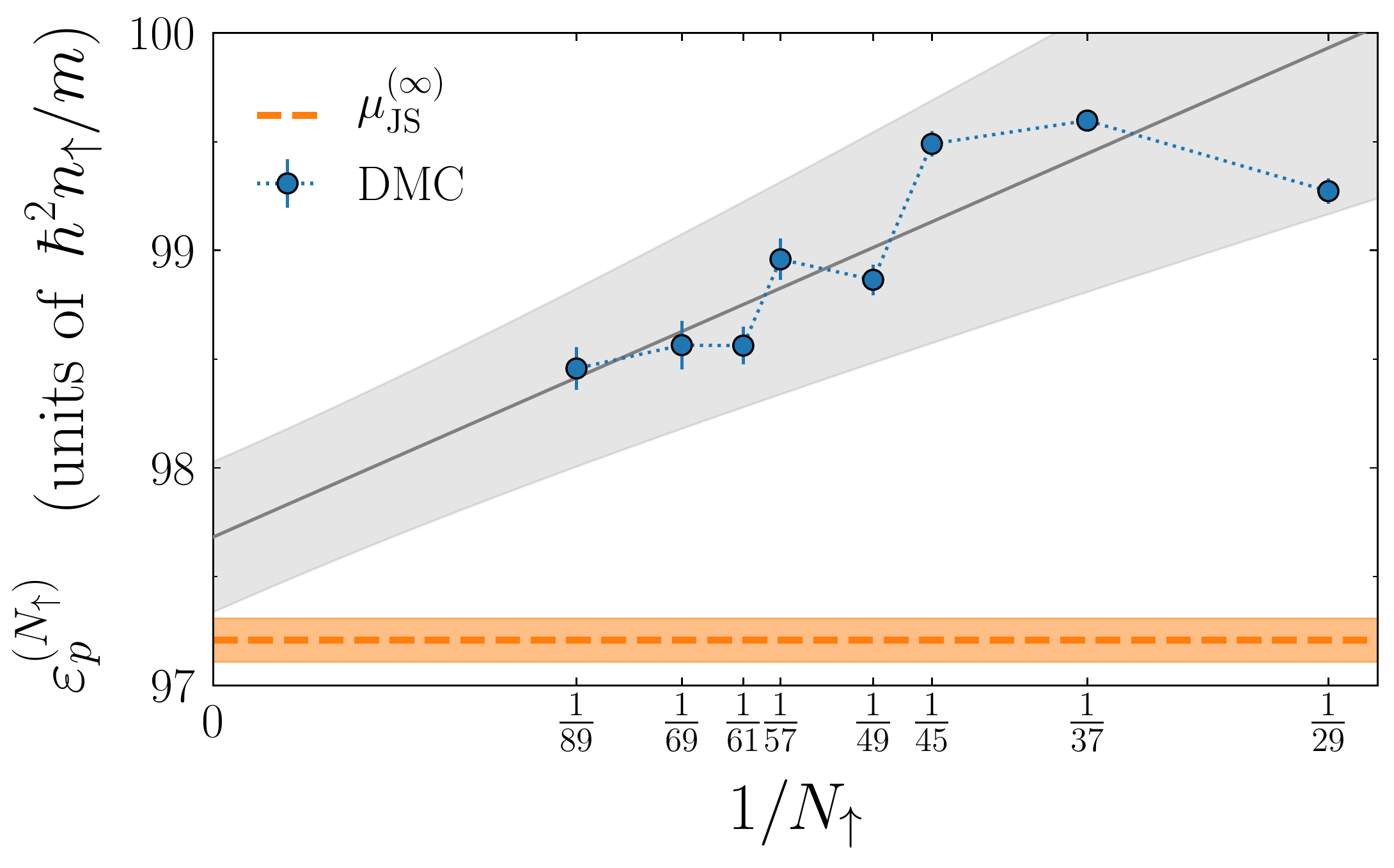}
\caption{Size scaling of the DMC polaron energy at density $n_\uparrow
r_0^2=40$, in units of $\hbar^2 n_\uparrow / m$ (blue circles -- blue dotted
line is a guide to the eye). For the linear fit (gray solid line), the shaded
band includes both the statistical uncertainty and the systematic error due to
the choice of the fit range. The $P = 1$ chemical potential
$\mu_\mathrm{JS}^{(\infty)}$ (orange horizontal dashed line, with shaded area
representing its uncertainty) is extracted from the equation of state -- see
text. All calculations are based on the $\Psi_\mathrm{JS}$ trial wave function.}
\label{fig:polaron}
\end{figure}

To be in the region where the ground state may have polarization $P = 1$, we
consider the density $n_\uparrow r_0^2 = 40$ (\emph{cf}.
Fig.~\ref{fig:DE_P0_P1}, and notice that $n = n_\uparrow$ in the thermodynamic
limit), where we perform calculations with the $\Psi_\mathrm{JS}$ trial wave
function. The polaron energy has a strong dependence on the system size, so
that a finite-size-scaling study is necessary. We find that DMC results are in
reasonable agreement with a linear scaling, $\varepsilon_p^{(N_\uparrow)} =
\varepsilon_p^{(\infty)} + \beta / N_\uparrow$ (see Fig.~\ref{fig:polaron}).
The best-fit result is $\varepsilon_p^{(\infty)} / (\hbar^2 n_\uparrow / m) =
97.7(3)$.

The chemical potential can be computed directly from the equation of state, as
\begin{equation}
\mu^{(N_\uparrow)} =
\left( 1 + n_\uparrow \frac{\partial}{\partial n_\uparrow} \right)
E_{(N_\uparrow, 0)}.
\end{equation}
We apply this definition to the JS equation of state of a large system
(\emph{cf}. Table~\ref{tab:energies_JS_BF}). We fit the energy per particle with
the expression $A_{1/2} n_\uparrow^{1/2} + A_1 n_\uparrow + A_{5/4}
n_\uparrow^{5/4} + A_{3/2} n_\uparrow^{3/2}$ \cite{Astrakharchik2007PRL}, from
which we obtain $\mu_\mathrm{JS}^{(\infty)} / (\hbar^2 n_\uparrow / m) =
97.2(1)$, for $n_\uparrow r_0^2 = 40$. The error bar includes the statistical
uncertainty and a systematic error due to finite-size effects.

The result concerning the stability against a spin flip, namely the sign of
$\Delta E_\mathrm{flip}$, is affected
by strong statistical and systematic uncertainties. Since the polaron energy has
a sizable dependence on $N_\uparrow$ (see Fig.~\ref{fig:polaron}), convergence
would only be obtained for larger system sizes. Such calculations are hindered
by the fact that $\varepsilon_p^{(N_\uparrow)}$ is computed as the difference
between two extensive observables, which makes it challenging to reach large
system sizes while keeping the statistical uncertainties under control. An
alternative method consists in estimating the difference between the total
energy of two systems with the same number of particles and density, namely with
$(N_\uparrow, N_\downarrow)$ equal to $(N - 1, 1)$ and $(N, 0)$. This approach reduces
the size dependence of $\varepsilon_p^{(N_\uparrow)}$, and in the case of a
bosonic bath it was successfully combined with the correlated-sampling
technique \cite{Boronat1999PRB}. For fermions, however, it introduces an
additional correction due to the change in the nodal structure of the $\uparrow$
component, and we found that it reaches a precision similar to the one reported
in this work.

In conclusion, we find that $\Delta E_\mathrm{flip} > 0$ at density $n r_0^2 =
40$, pointing towards the stability of the $P = 1$ state. However, the large
uncertainties on $\Delta E_\mathrm{flip}$ make this result inconclusive, and we
cannot use it as a reliable confirmation of the fully-polarized ground state
identified in Section~\ref{sect:ferromagnetism:direct} for densities higher than
$n r_0^2 \approx 26$.

\subsection{Iterative-backflow wave functions}
\label{sect:ferromagnetism:iterativeBF}

As observed above, the fixed-node bias cannot be neglected in the study of
itinerant ferromagnetism. In this section we investigate this issue in detail,
by making use of improved trial wave functions based on the iterative backflow
transformations \cite{Taddei2015PRB}.

\begin{table*}[tb]
\begin{center}
\begin{tabular}{c | c c c | c c c | c c}
& & $P = 1$ &
& & $P = 0$ &
& $\Delta E_T$
& $\Delta E$ \\

$\Psi_T$ &
$\sigma_T^2$ & $E_T$ & $E$ &
$\sigma_T^2$ & $E_T$ & $E$ \\
\hline
JS        & 102(8)   & 1615.98(5) & 1607.92(1) & 311(10)  & 1636.56(8)  & 1610.85(5)  & 20.6(1)  & 2.93(6) \\
JS3       &  47(2)   & 1610.90(3) & - &          185(6)   & 1625.21(6)  & -           & 14.31(6) & - \\
BF        &  40(2)   & 1609.28(3) & 1607.09(2) &  98(2)   & 1613.94(17) & 1606.91(17) & 4.7(2)   & -0.2(2)\\
BF3       &  28.5(8) & 1608.59(7) & 1606.94(4) &  96(2)   & 1612.63(7)  & 1606.29(6)  & 4.0(1)   & -0.65(7) \\
IT1       &  32.6(7) & 1608.40(7) & 1606.91(3) &  78(1)   & 1610.78(11) & 1605.69(9)  & 2.38(13) & -1.22(1) \\
BF3-IT1   &  26.8(5) & 1608.15(7) & 1606.87(5) &  75(2)   & 1609.65(11) & 1605.57(6)  & 1.50(13) & -1.31(8) \\
IT2       &  30.6(7) & 1608.02(6) & 1606.91(6)  & 74.8(9) & 1609.54(7)  & 1605.31(13) & 1.52(9)  & -1.59(14) \\
VMC$_\mathrm{ext}$ & & 1605.3(5)  &            &          & 1602(1)     &             &          & \\
\end{tabular}
\end{center}
\caption{Variance and energy per particle at density $n r_0^2 = 40$, in units of
$\varepsilon_0$, for different trial wave functions $\Psi_T$. The variational
energy per particle $E_T$ and the DMC result $E$ are reported, and for $E_T$ we
also report the variance $\sigma_T^2$. The last two columns report the energy
difference between the unpolarized and polarized states, for variational results
[$\Delta E_T \equiv E_{T, P=0} - E_{T, P=1}$] and for DMC energies [$\Delta E$,
\emph{cf}. Eq.~\eqref{eq:DeltaE}]. The line ``VMC$_\mathrm{ext}$'' is obtained
through a linear extrapolation of the $E_T$ values -- see text. Data for $P=1$
($P=0$) are obtained with $N = 121$ ($N = 122$) particles. For the details about
the wave functions, and for a comparison with Table~\ref{tab:energies_JS_BF},
see Appendix~\ref{app:iterative_backflow}. Note that the JS and JS3 wave
functions have the same nodal structure, so that their DMC energies should be
the same within statistical uncertainty.
}
\label{tab:iterativeBF}
\end{table*}

Given a variational ansatz $\Psi_T$, the energy per particle $E_T$
and its variance $\sigma_T^2$ are defined as
\begin{align}
E_T & \equiv
\frac{\left \langle \Psi_T \right| H \left | \Psi_T \right \rangle}{N}, \\
\sigma^2_T & \equiv
\frac{\left \langle \Psi_T \right| \left( H - N E_T \right)^2 \left | \Psi_T \right \rangle} {N^2}.
\end{align}
These two observables are related to the quality of the ansatz
\cite{Mora2007PRL}: When $\Psi_T$ approaches an eigenstate $\Psi_0$ of $H$,
$E_T$ tends to $E_0 \equiv \langle \Psi_0 | H | \Psi_0 \rangle / N$ and
$\sigma^2_T$ tends to zero. Rigorous bounds connect $E_T$ to
$E_0$ and $\sigma_T^2$ \cite{Temple1928PRSL, Weinstein1934PNAS, Goedecker1991PRB}. We assume that the eigenstate
$\Psi_0$ closest in energy to $\Psi_T$ is the ground state, so that $E_0$ is the
ground-state energy per particle. We also assume that $\langle \Psi_T | \Psi_0
\rangle$ is the only significant overlap of $\Psi_T$ with an eigenstate of $H$,
which is reasonable for an extended system. In this case, we obtain a linear
relation between $\sigma_T^2$ and $E_T$ \cite{Taddei2015PRB},
\begin{equation}
E_T = E_0 + N A \sigma_T^2 \qquad \text{for $\sigma_T^2\to0$},
\label{eq:energy_variance_extrapolation}
\end{equation}
where $A$ is a constant coefficient.

Eq.~\eqref{eq:energy_variance_extrapolation} provides a practical way to
estimate the true ground-state energy $E_0$, by computing $\sigma^2_T$ and $E_T$
for several trial wave functions (through the Variational Monte Carlo method)
and performing a linear extrapolation towards $\sigma_T^2=0$. The validity of
this extrapolation was verified for both $^3$He and $^4$He \cite{Taddei2015PRB,
Ruggeri2018PRL}. For $^3$He, the technique can be benchmarked against independent
calculations of the ground-state energy, obtained through the transient-estimate
method for polarized and unpolarized systems. The result is that the
energy-variance extrapolated energy has an accuracy comparable with the
best available DMC results \cite{Taddei2015PRB}.

\begin{figure}[htb]
\centering
\includegraphics[width=\linewidth]{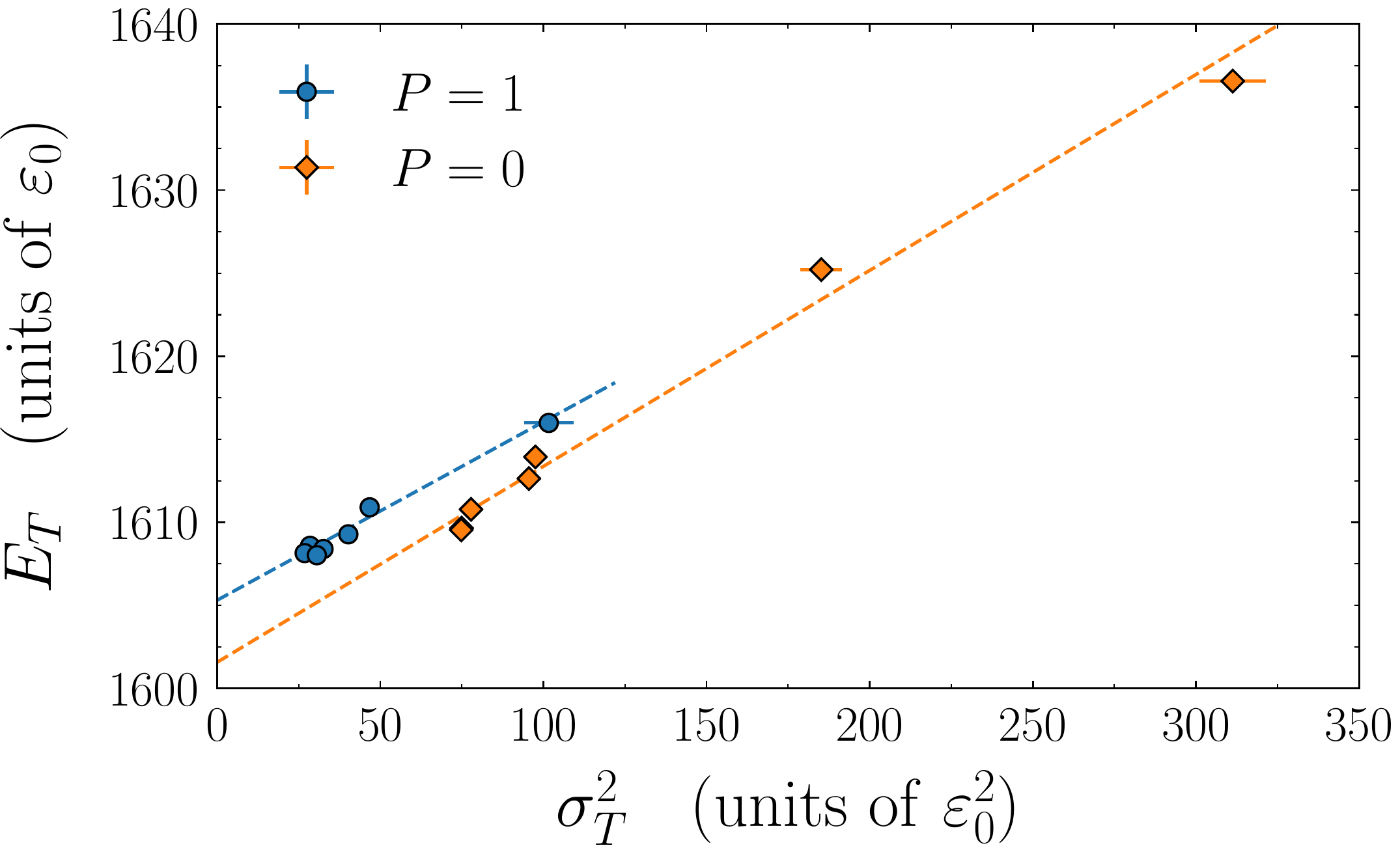}
\caption{Scaling of the energy per particle \emph{vs} its variance, for different
variational wave functions, at polarizations $P = 1$ (blue circles) and $P = 0$
(orange diamonds) -- \emph{cf.} Table~\ref{tab:iterativeBF}. Dashed lines are
linear fits to the data, see Eq.~\eqref{eq:energy_variance_extrapolation}.}
\label{fig:iterativeBF}
\end{figure}

We applied this extrapolation scheme to the variational energies of the
polarized and unpolarized dipolar Fermi gas at density $n r_0^2 = 40$, as shown
in Fig.~\ref{fig:iterativeBF}. For any choice of $\Psi_T$, the energy difference
$\Delta E_T \equiv E_{T, P=0} - E_{T, P=1}$ is positive (see 8th column in
Table~\ref{tab:iterativeBF}), indicating that the $P
= 1$ state is always favored at the variational level, for the wave functions
considered here. Nevertheless, the energy gain in using a better trial wave
function is more relevant for the $P = 0$ state (\emph{cf}. $E_T$ columns in
Table~\ref{tab:iterativeBF}), as noted in
Section~\ref{sect:ferromagnetism:direct}. As a consequence, the energy
difference $\Delta E_T$ decreases upon improving the trial wave function. We
fit the data for $E_T$ with Eq.~\eqref{eq:energy_variance_extrapolation},
finding a reasonable linear scaling. We then estimate the true ground-state
energy through the intercept with the vertical axis, and we find that the lowest
energy corresponds to the $P = 0$ state. We conclude that the fully-polarized
state is not the ground state at density $n r_0^2 = 40$.

An independent confirmation comes from the fixed-node DMC energies for
simulations based on the iterative-backflow wave functions -- see
Table~\ref{tab:iterativeBF}. For $P = 1$, our four best choices for the nodal
structure yield energies which are compatible with each other within statistical
uncertainties. For $P = 0$, the energy dependence on the wave function is
larger, and the lowest energy is reached through the iterative-backflow wave
function IT2.
As reported in the last column of Table~\ref{tab:iterativeBF}, and at a
difference with the case of variational energies, the DMC energy difference
$\Delta E$ becomes negative when the accuracy of
the nodal structure increases. The best available DMC energies are indeed lower
for the $P = 0$ state, in agreement with the energy-variance extrapolation.

The evidence for the fact that the ground state is not fully polarized extends
to the density $n r_0^2 = 48$ \cite{Comparin2018Zenodo}. In that case, the best
DMC energy for
$P = 1$, $E_\mathrm{BF3-IT1} = 2081.69(5)$, is still higher than the one for
$P=0$, $E_\mathrm{BF3-IT1} = 2080.49(15)$.
Also the energy-variance extrapolation leads to the same conclusion. Therefore, we
exclude the existence of a fully-polarized ground state for densities up to $n
r_0^2 = 48$.

\section{Discussion and conclusions}

We performed the first study of the $P = 0$ mixture of dipolar fermions, both at
low and high density, computing the equation of state and pair distribution
functions through the DMC technique. For density below $n r_0^2 \approx
10^{-3}$, the equation of state agrees within a few percent both with a
mean-field approximation and with the energy of fermionic hard disks.
Non-universal beyond-mean-field corrections are present for higher densities.

Two assumptions of the theoretical model need to be discussed, in view of a
connection to experiments with ultracold atoms or molecules: The reduced
dimensionality and the shape of the interparticle potential. The experimental
realization of a two-dimensional system is based on a tight confinement along
the transverse direction, characterized by the harmonic-oscillator length $a_z$ or by
the typical energy $\hbar \omega_z$. At zero temperature, the condition to be in
the two-dimensional regime reads $\mu \ll \hbar \omega_z$, where $\mu$ is the
chemical potential, which is of the order of
$E_\mathrm{IFG}$. Such condition corresponds to
$n r_0^2 \ll ({r_0}/{a_z})^2$,
showing that the maximum allowed value of $n r_0^2$ depends on the ratio $a_z /
r_0$. By setting $r_0 \approx 20$ nm (the value for Dy atoms) and choosing a
realistic value $a_z \approx 500$ nm, we find that the confined system can be
described by a two-dimensional model up to $n r_0^2 \approx 10^{-3}$. This
density is comparable with the one where beyond-mean-field corrections to
the equation of state become important.

The second issue is that our model neglects the presence of an additional contact interaction,
on top of the dipolar repulsion. On the one hand, the two-dimensional scattering
length for a three-dimensional contact interaction with scattering length
$a_{3D}$ scales as $\exp(- \sqrt{\frac{\pi}{2}} a_z / a_{3D})$, in presence of
transverse confinement \cite{Pitaevskii2016}. Thus it is strongly suppressed
when $a_{3D} \ll a_z$, which is the typical case away from Feshbach resonances.
On the other hand, the two-dimensional scattering length of the dipolar
potential is of
the order of $r_0$, independently on $a_z$. For this reason, the lack of a
contact interaction in our model should not have a relevant effect on the
equation of state, at the experimentally realistic densities $n r_0^2 = 10^{-3}
- 10^{-1}$.

When moving towards high values of $n r_0^2$, the aforementioned condition for
being in the two-dimensional regime implies that a large value of $r_0 / a_z$ is
needed. This is currently beyond reach for the permanent magnetic moment of
atoms, but could be achieved for a gas of molecules with a large induced dipole
moment -- see for instance Refs.~\cite{Ni2010, Park2015PRL}.

At high density, we find that the issue of itinerant ferromagnetism is extremely
subtle due to the small energy differences at play. Thus it requires a high
degree of accuracy in the nodal structure, to be treated within the fixed-node
scheme. In our analysis
we observe no signature of a
fully-polarized ground state, up to a density approximately equal to
the freezing density. Once the freezing transition is crossed, we expect a weak
dependence of the energy on the polarization, in analogy with the case of the
two-dimensional electron gas \cite{Drummond2009PRL}. In principle there may
exist an intermediate-density regime where the ground state is partially
polarized, with $0 < P < 1$, as in the case of three-dimensional hard-core
fermions \cite{Pilati2010PRL}. This possibility remains open, as our current
study of dipolar fermions is restricted to polarization $P = 0$ and $1$.

We also demonstrate the usefulness of iterative-backflow wave functions in a novel
system, where the accuracy of the common JS/BF wave functions is inadequate for
the study of itinerant ferromagnetism. The combination of high-accuracy DMC
results with the energy-variance extrapolation of variational data provides a
clear indication about the energies of the polarized and unpolarized states, and
meanwhile constitutes an estimate of the fixed-node bias for the two cases.

\acknowledgments{We acknowledge Luca Parisi for useful discussions, and Gianluca
Bertaina for comments and for providing the data from
Ref.~\cite{Bertaina2013EPJ}. This work has been supported by the Provincia
Autonoma di Trento and by the Ministerio de Economia, Industria y Competitividad
(ES) under Grant No.  FIS2017-84114-C2-1-P and FPI fellowship BES2015-074088.
T. C. thanks the Institute for Nuclear Theory (University of Washington) for
hospitality. Data and additional details about the numerical simulations are
made publicly available \cite{Comparin2018Zenodo}.}

\appendix

\section{Trial wave functions for Section ~\ref{sect:ferromagnetism:iterativeBF}}
\label{app:iterative_backflow}

In this appendix, we describe the trial wave functions used in
Section~\ref{sect:ferromagnetism:iterativeBF}. The corresponding numerical
results are obtained through an independent implementation of the DMC algorithm,
with respect to the rest of the manuscript.

The JS and JS3 wave functions are of the standard Jastrow-Slater form, with
two-body (JS) or two- and three-body (JS3) Jastrow correlations
\cite{Holzmann2006PRB}. Each Jastrow correlation function is parametrized via
locally piecewise--quintic Hermite interpolants (splines) with typically 38
variational parameters per function \cite{Natoli1995JCP, Holzmann2005JCP}.
Although the variational energies $E_T$ of the JS and JS3 wave function are
different (see Table~\ref{tab:iterativeBF}), the two DMC energies have to be the
same since the two wave functions share the same nodal structure. Moreover, the
JS results of our two independent implementations of the DMC algorithm agree
within error bars
(compare Table~\ref{tab:energies_JS_BF} with
Table~\ref{tab:iterativeBF}, for density $n r_0^2 = 40$), which constitutes an
additional validity check.

The BF wave function includes two-body backflow correlations as in
Eq.~\eqref{eq:BF}, with $f_\mathrm{BF}(r)$ parametrized through Hermite
interpolants (as in the case of Jastrow correlations). Different choices of
$f_\mathrm{BF}(r)$ correspond to different nodal structures, so that also the
DMC energy can differ. As an example, the DMC energy for the BF wave function
parametrized as in Ref.~\cite{Grau2002PRL} is larger than the one where
$f_\mathrm{BF}(r)$ is parametrized through splines -- \emph{cf}. ``BF'' entries
in Table~\ref{tab:energies_JS_BF} (for $n r_0^2 = 40$) and in
Table~\ref{tab:iterativeBF}. Note that this does not affect the conclusions of
this work. In the BF3 case, also an explicit three-body backflow correction is
included, as in Ref.~\cite{Holzmann2006PRB}.

As a starting point of the iterative backflow procedure, we choose the IT0 wave
function (that is, the one with $\alpha = 0$) to be the Jastrow-Slater wave
function with two-body Jastrow correlations. At each iteration $\alpha \to
\alpha + 1$, we perform an additional transformation from
$\mathbf{q}_i^{(\alpha)}$ to
\begin{equation}
\mathbf{q}_i^{(\alpha + 1)}
\equiv
\mathbf{q}_i^{(\alpha)}
+
\sum_{j\neq i} \left(\mathbf{q}_i^{(\alpha)} - \mathbf{q}_j^{(\alpha)} \right)
f_\mathrm{BF}^{(\alpha)}\left( | \mathbf{q}_i^{(\alpha)} -
\mathbf{q}_j^{(\alpha)} | \right),
\end{equation}
where $\mathbf{q}_i^{(0)} = \mathbf{x}_i$ are the particle coordinates.
This procedure leads to the IT1 and IT2 wave functions.
The iterative-backflow transformations $f_\mathrm{BF}^{(\alpha)}(r)$ are
parametrized via three-parameters Gaussian functions, $A \exp\left[- B (r - C)^2
\right]$,
additionally imposing that the function and its derivative vanish at $r = L/2$.
The ``bare'' Jastrow and backflow potentials are still treated
via Hermite interpolants.

For the BF3IT1 wave function we combine the non-iterated three-body backflow
correlation (as in the BF3 wave function) with an iterated two-body backflow and
Jastrow potential (as in the IT1 wave function).

\bibliography{refs}

\begin{thebibliography}{82}%
\makeatletter
\providecommand \@ifxundefined [1]{%
 \@ifx{#1\undefined}
}%
\providecommand \@ifnum [1]{%
 \ifnum #1\expandafter \@firstoftwo
 \else \expandafter \@secondoftwo
 \fi
}%
\providecommand \@ifx [1]{%
 \ifx #1\expandafter \@firstoftwo
 \else \expandafter \@secondoftwo
 \fi
}%
\providecommand \natexlab [1]{#1}%
\providecommand \enquote  [1]{``#1''}%
\providecommand \bibnamefont  [1]{#1}%
\providecommand \bibfnamefont [1]{#1}%
\providecommand \citenamefont [1]{#1}%
\providecommand \href@noop [0]{\@secondoftwo}%
\providecommand \href [0]{\begingroup \@sanitize@url \@href}%
\providecommand \@href[1]{\@@startlink{#1}\@@href}%
\providecommand \@@href[1]{\endgroup#1\@@endlink}%
\providecommand \@sanitize@url [0]{\catcode `\\12\catcode `\$12\catcode
  `\&12\catcode `\#12\catcode `\^12\catcode `\_12\catcode `\%12\relax}%
\providecommand \@@startlink[1]{}%
\providecommand \@@endlink[0]{}%
\providecommand \url  [0]{\begingroup\@sanitize@url \@url }%
\providecommand \@url [1]{\endgroup\@href {#1}{\urlprefix }}%
\providecommand \urlprefix  [0]{URL }%
\providecommand \Eprint [0]{\href }%
\providecommand \doibase [0]{http://dx.doi.org/}%
\providecommand \selectlanguage [0]{\@gobble}%
\providecommand \bibinfo  [0]{\@secondoftwo}%
\providecommand \bibfield  [0]{\@secondoftwo}%
\providecommand \translation [1]{[#1]}%
\providecommand \BibitemOpen [0]{}%
\providecommand \bibitemStop [0]{}%
\providecommand \bibitemNoStop [0]{.\EOS\space}%
\providecommand \EOS [0]{\spacefactor3000\relax}%
\providecommand \BibitemShut  [1]{\csname bibitem#1\endcsname}%
\let\auto@bib@innerbib\@empty
\bibitem [{\citenamefont {Dalfovo}\ \emph {et~al.}(1999)\citenamefont
  {Dalfovo}, \citenamefont {Giorgini}, \citenamefont {Pitaevskii},\ and\
  \citenamefont {Stringari}}]{Dalfovo1999RMP}%
  \BibitemOpen
  \bibfield  {author} {\bibinfo {author} {\bibfnamefont {F.}~\bibnamefont
  {Dalfovo}}, \bibinfo {author} {\bibfnamefont {S.}~\bibnamefont {Giorgini}},
  \bibinfo {author} {\bibfnamefont {L.~P.}\ \bibnamefont {Pitaevskii}}, \ and\
  \bibinfo {author} {\bibfnamefont {S.}~\bibnamefont {Stringari}},\ }\href
  {\doibase 10.1103/RevModPhys.71.463} {\bibfield  {journal} {\bibinfo
  {journal} {Rev. Mod. Phys.}\ }\textbf {\bibinfo {volume} {71}},\ \bibinfo
  {pages} {463} (\bibinfo {year} {1999})}\BibitemShut {NoStop}%
\bibitem [{\citenamefont {Giorgini}\ \emph {et~al.}(2008)\citenamefont
  {Giorgini}, \citenamefont {Pitaevskii},\ and\ \citenamefont
  {Stringari}}]{Giorgini2008RMP}%
  \BibitemOpen
  \bibfield  {author} {\bibinfo {author} {\bibfnamefont {S.}~\bibnamefont
  {Giorgini}}, \bibinfo {author} {\bibfnamefont {L.~P.}\ \bibnamefont
  {Pitaevskii}}, \ and\ \bibinfo {author} {\bibfnamefont {S.}~\bibnamefont
  {Stringari}},\ }\href {\doibase 10.1103/RevModPhys.80.1215} {\bibfield
  {journal} {\bibinfo  {journal} {Rev. Mod. Phys.}\ }\textbf {\bibinfo {volume}
  {80}},\ \bibinfo {pages} {1215} (\bibinfo {year} {2008})}\BibitemShut
  {NoStop}%
\bibitem [{\citenamefont {Baranov}(2008)}]{Baranov2008PR}%
  \BibitemOpen
  \bibfield  {author} {\bibinfo {author} {\bibfnamefont {M.~A.}\ \bibnamefont
  {Baranov}},\ }\href {\doibase https://doi.org/10.1016/j.physrep.2008.04.007}
  {\bibfield  {journal} {\bibinfo  {journal} {Phys. Rep.}\ }\textbf {\bibinfo
  {volume} {464}},\ \bibinfo {pages} {71 } (\bibinfo {year}
  {2008})}\BibitemShut {NoStop}%
\bibitem [{\citenamefont {Lahaye}\ \emph {et~al.}(2009)\citenamefont {Lahaye},
  \citenamefont {Menotti}, \citenamefont {Santos}, \citenamefont {Lewenstein},\
  and\ \citenamefont {Pfau}}]{Lahaye2009RPP}%
  \BibitemOpen
  \bibfield  {author} {\bibinfo {author} {\bibfnamefont {T.}~\bibnamefont
  {Lahaye}}, \bibinfo {author} {\bibfnamefont {C.}~\bibnamefont {Menotti}},
  \bibinfo {author} {\bibfnamefont {L.}~\bibnamefont {Santos}}, \bibinfo
  {author} {\bibfnamefont {M.}~\bibnamefont {Lewenstein}}, \ and\ \bibinfo
  {author} {\bibfnamefont {T.}~\bibnamefont {Pfau}},\ }\href {\doibase
  10.1088/0034-4885/72/12/126401} {\bibfield  {journal} {\bibinfo  {journal}
  {Rep. Progr. Phys.}\ }\textbf {\bibinfo {volume} {72}},\ \bibinfo {pages}
  {126401} (\bibinfo {year} {2009})}\BibitemShut {NoStop}%
\bibitem [{\citenamefont {Bohn}\ \emph {et~al.}(2017)\citenamefont {Bohn},
  \citenamefont {Rey},\ and\ \citenamefont {Ye}}]{Bohn2017}%
  \BibitemOpen
  \bibfield  {author} {\bibinfo {author} {\bibfnamefont {J.~L.}\ \bibnamefont
  {Bohn}}, \bibinfo {author} {\bibfnamefont {A.~M.}\ \bibnamefont {Rey}}, \
  and\ \bibinfo {author} {\bibfnamefont {J.}~\bibnamefont {Ye}},\ }\href
  {\doibase 10.1126/science.aam6299} {\bibfield  {journal} {\bibinfo  {journal}
  {Science}\ }\textbf {\bibinfo {volume} {357}},\ \bibinfo {pages} {1002}
  (\bibinfo {year} {2017})}\BibitemShut {NoStop}%
\bibitem [{\citenamefont {Santos}\ \emph {et~al.}(2000)\citenamefont {Santos},
  \citenamefont {Shlyapnikov}, \citenamefont {Zoller},\ and\ \citenamefont
  {Lewenstein}}]{Santos2000PRL}%
  \BibitemOpen
  \bibfield  {author} {\bibinfo {author} {\bibfnamefont {L.}~\bibnamefont
  {Santos}}, \bibinfo {author} {\bibfnamefont {G.~V.}\ \bibnamefont
  {Shlyapnikov}}, \bibinfo {author} {\bibfnamefont {P.}~\bibnamefont {Zoller}},
  \ and\ \bibinfo {author} {\bibfnamefont {M.}~\bibnamefont {Lewenstein}},\
  }\href {\doibase 10.1103/PhysRevLett.85.1791} {\bibfield  {journal} {\bibinfo
   {journal} {Phys. Rev. Lett.}\ }\textbf {\bibinfo {volume} {85}},\ \bibinfo
  {pages} {1791} (\bibinfo {year} {2000})}\BibitemShut {NoStop}%
\bibitem [{\citenamefont {Aikawa}\ \emph {et~al.}(2014)\citenamefont {Aikawa},
  \citenamefont {Baier}, \citenamefont {Frisch}, \citenamefont {Mark},
  \citenamefont {Ravensbergen},\ and\ \citenamefont {Ferlaino}}]{Aikawa2014}%
  \BibitemOpen
  \bibfield  {author} {\bibinfo {author} {\bibfnamefont {K.}~\bibnamefont
  {Aikawa}}, \bibinfo {author} {\bibfnamefont {S.}~\bibnamefont {Baier}},
  \bibinfo {author} {\bibfnamefont {A.}~\bibnamefont {Frisch}}, \bibinfo
  {author} {\bibfnamefont {M.}~\bibnamefont {Mark}}, \bibinfo {author}
  {\bibfnamefont {C.}~\bibnamefont {Ravensbergen}}, \ and\ \bibinfo {author}
  {\bibfnamefont {F.}~\bibnamefont {Ferlaino}},\ }\href {\doibase
  10.1126/science.1255259} {\bibfield  {journal} {\bibinfo  {journal}
  {Science}\ }\textbf {\bibinfo {volume} {345}},\ \bibinfo {pages} {1484}
  (\bibinfo {year} {2014})}\BibitemShut {NoStop}%
\bibitem [{\citenamefont {Baier}\ \emph {et~al.}(2016)\citenamefont {Baier},
  \citenamefont {Mark}, \citenamefont {Petter}, \citenamefont {Aikawa},
  \citenamefont {Chomaz}, \citenamefont {Cai}, \citenamefont {Baranov},
  \citenamefont {Zoller},\ and\ \citenamefont {Ferlaino}}]{Baier2016}%
  \BibitemOpen
  \bibfield  {author} {\bibinfo {author} {\bibfnamefont {S.}~\bibnamefont
  {Baier}}, \bibinfo {author} {\bibfnamefont {M.~J.}\ \bibnamefont {Mark}},
  \bibinfo {author} {\bibfnamefont {D.}~\bibnamefont {Petter}}, \bibinfo
  {author} {\bibfnamefont {K.}~\bibnamefont {Aikawa}}, \bibinfo {author}
  {\bibfnamefont {L.}~\bibnamefont {Chomaz}}, \bibinfo {author} {\bibfnamefont
  {Z.}~\bibnamefont {Cai}}, \bibinfo {author} {\bibfnamefont {M.}~\bibnamefont
  {Baranov}}, \bibinfo {author} {\bibfnamefont {P.}~\bibnamefont {Zoller}}, \
  and\ \bibinfo {author} {\bibfnamefont {F.}~\bibnamefont {Ferlaino}},\ }\href
  {\doibase 10.1126/science.aac9812} {\bibfield  {journal} {\bibinfo  {journal}
  {Science}\ }\textbf {\bibinfo {volume} {352}},\ \bibinfo {pages} {201}
  (\bibinfo {year} {2016})}\BibitemShut {NoStop}%
\bibitem [{\citenamefont {Chomaz}\ \emph {et~al.}(2016)\citenamefont {Chomaz},
  \citenamefont {Baier}, \citenamefont {Petter}, \citenamefont {Mark},
  \citenamefont {W\"{a}chtler}, \citenamefont {Santos},\ and\ \citenamefont
  {Ferlaino}}]{Chomaz2016PRX}%
  \BibitemOpen
  \bibfield  {author} {\bibinfo {author} {\bibfnamefont {L.}~\bibnamefont
  {Chomaz}}, \bibinfo {author} {\bibfnamefont {S.}~\bibnamefont {Baier}},
  \bibinfo {author} {\bibfnamefont {D.}~\bibnamefont {Petter}}, \bibinfo
  {author} {\bibfnamefont {M.~J.}\ \bibnamefont {Mark}}, \bibinfo {author}
  {\bibfnamefont {F.}~\bibnamefont {W\"{a}chtler}}, \bibinfo {author}
  {\bibfnamefont {L.}~\bibnamefont {Santos}}, \ and\ \bibinfo {author}
  {\bibfnamefont {F.}~\bibnamefont {Ferlaino}},\ }\href {\doibase
  10.1103/PhysRevX.6.041039} {\bibfield  {journal} {\bibinfo  {journal} {Phys.
  Rev. X}\ }\textbf {\bibinfo {volume} {6}},\ \bibinfo {pages} {041039}
  (\bibinfo {year} {2016})}\BibitemShut {NoStop}%
\bibitem [{\citenamefont {Ferrier-Barbut}\ \emph {et~al.}(2016)\citenamefont
  {Ferrier-Barbut}, \citenamefont {Kadau}, \citenamefont {Schmitt},
  \citenamefont {Wenzel},\ and\ \citenamefont {Pfau}}]{FerrierBarbut2016PRL}%
  \BibitemOpen
  \bibfield  {author} {\bibinfo {author} {\bibfnamefont {I.}~\bibnamefont
  {Ferrier-Barbut}}, \bibinfo {author} {\bibfnamefont {H.}~\bibnamefont
  {Kadau}}, \bibinfo {author} {\bibfnamefont {M.}~\bibnamefont {Schmitt}},
  \bibinfo {author} {\bibfnamefont {M.}~\bibnamefont {Wenzel}}, \ and\ \bibinfo
  {author} {\bibfnamefont {T.}~\bibnamefont {Pfau}},\ }\href {\doibase
  10.1103/PhysRevLett.116.215301} {\bibfield  {journal} {\bibinfo  {journal}
  {Phys. Rev. Lett.}\ }\textbf {\bibinfo {volume} {116}},\ \bibinfo {pages}
  {215301} (\bibinfo {year} {2016})}\BibitemShut {NoStop}%
\bibitem [{\citenamefont {Schmitt}\ \emph {et~al.}(2016)\citenamefont
  {Schmitt}, \citenamefont {Wenzel}, \citenamefont {B\"{o}ttcher},
  \citenamefont {Ferrier-Barbut},\ and\ \citenamefont {Pfau}}]{Schmitt2016}%
  \BibitemOpen
  \bibfield  {author} {\bibinfo {author} {\bibfnamefont {M.}~\bibnamefont
  {Schmitt}}, \bibinfo {author} {\bibfnamefont {M.}~\bibnamefont {Wenzel}},
  \bibinfo {author} {\bibfnamefont {F.}~\bibnamefont {B\"{o}ttcher}}, \bibinfo
  {author} {\bibfnamefont {I.}~\bibnamefont {Ferrier-Barbut}}, \ and\ \bibinfo
  {author} {\bibfnamefont {T.}~\bibnamefont {Pfau}},\ }\href {\doibase
  10.1038/nature20126} {\bibfield  {journal} {\bibinfo  {journal} {Nature}\
  }\textbf {\bibinfo {volume} {539}},\ \bibinfo {pages} {259} (\bibinfo {year}
  {2016})}\BibitemShut {NoStop}%
\bibitem [{\citenamefont {Baier}\ \emph {et~al.}(2018)\citenamefont {Baier},
  \citenamefont {Petter}, \citenamefont {Becher}, \citenamefont {Patscheider},
  \citenamefont {Natale}, \citenamefont {Chomaz}, \citenamefont {Mark},\ and\
  \citenamefont {Ferlaino}}]{Baier2018PRL}%
  \BibitemOpen
  \bibfield  {author} {\bibinfo {author} {\bibfnamefont {S.}~\bibnamefont
  {Baier}}, \bibinfo {author} {\bibfnamefont {D.}~\bibnamefont {Petter}},
  \bibinfo {author} {\bibfnamefont {J.~H.}\ \bibnamefont {Becher}}, \bibinfo
  {author} {\bibfnamefont {A.}~\bibnamefont {Patscheider}}, \bibinfo {author}
  {\bibfnamefont {G.}~\bibnamefont {Natale}}, \bibinfo {author} {\bibfnamefont
  {L.}~\bibnamefont {Chomaz}}, \bibinfo {author} {\bibfnamefont {M.~J.}\
  \bibnamefont {Mark}}, \ and\ \bibinfo {author} {\bibfnamefont
  {F.}~\bibnamefont {Ferlaino}},\ }\href {\doibase
  10.1103/PhysRevLett.121.093602} {\bibfield  {journal} {\bibinfo  {journal}
  {Phys. Rev. Lett.}\ }\textbf {\bibinfo {volume} {121}},\ \bibinfo {pages}
  {093602} (\bibinfo {year} {2018})}\BibitemShut {NoStop}%
\bibitem [{\citenamefont {Trautmann}\ \emph {et~al.}(2018)\citenamefont
  {Trautmann}, \citenamefont {Ilzh\"ofer}, \citenamefont {Durastante},
  \citenamefont {Politi}, \citenamefont {Sohmen}, \citenamefont {Mark},\ and\
  \citenamefont {Ferlaino}}]{Trautmann2018PRL}%
  \BibitemOpen
  \bibfield  {author} {\bibinfo {author} {\bibfnamefont {A.}~\bibnamefont
  {Trautmann}}, \bibinfo {author} {\bibfnamefont {P.}~\bibnamefont
  {Ilzh\"ofer}}, \bibinfo {author} {\bibfnamefont {G.}~\bibnamefont
  {Durastante}}, \bibinfo {author} {\bibfnamefont {C.}~\bibnamefont {Politi}},
  \bibinfo {author} {\bibfnamefont {M.}~\bibnamefont {Sohmen}}, \bibinfo
  {author} {\bibfnamefont {M.~J.}\ \bibnamefont {Mark}}, \ and\ \bibinfo
  {author} {\bibfnamefont {F.}~\bibnamefont {Ferlaino}},\ }\href {\doibase
  10.1103/PhysRevLett.121.213601} {\bibfield  {journal} {\bibinfo  {journal}
  {Phys. Rev. Lett.}\ }\textbf {\bibinfo {volume} {121}},\ \bibinfo {pages}
  {213601} (\bibinfo {year} {2018})}\BibitemShut {NoStop}%
\bibitem [{\citenamefont {Partridge}\ \emph {et~al.}(2006)\citenamefont
  {Partridge}, \citenamefont {Li}, \citenamefont {Kamar}, \citenamefont
  {Liao},\ and\ \citenamefont {Hulet}}]{Partridge2006}%
  \BibitemOpen
  \bibfield  {author} {\bibinfo {author} {\bibfnamefont {G.~B.}\ \bibnamefont
  {Partridge}}, \bibinfo {author} {\bibfnamefont {W.}~\bibnamefont {Li}},
  \bibinfo {author} {\bibfnamefont {R.~I.}\ \bibnamefont {Kamar}}, \bibinfo
  {author} {\bibfnamefont {Y.-a.}\ \bibnamefont {Liao}}, \ and\ \bibinfo
  {author} {\bibfnamefont {R.~G.}\ \bibnamefont {Hulet}},\ }\href {\doibase
  10.1126/science.1122876} {\bibfield  {journal} {\bibinfo  {journal}
  {Science}\ }\textbf {\bibinfo {volume} {311}},\ \bibinfo {pages} {503}
  (\bibinfo {year} {2006})}\BibitemShut {NoStop}%
\bibitem [{\citenamefont {Stamper-Kurn}\ and\ \citenamefont
  {Ueda}(2013)}]{StamperKurn2013RMP}%
  \BibitemOpen
  \bibfield  {author} {\bibinfo {author} {\bibfnamefont {D.~M.}\ \bibnamefont
  {Stamper-Kurn}}\ and\ \bibinfo {author} {\bibfnamefont {M.}~\bibnamefont
  {Ueda}},\ }\href {\doibase 10.1103/RevModPhys.85.1191} {\bibfield  {journal}
  {\bibinfo  {journal} {Rev. Mod. Phys.}\ }\textbf {\bibinfo {volume} {85}},\
  \bibinfo {pages} {1191} (\bibinfo {year} {2013})}\BibitemShut {NoStop}%
\bibitem [{\citenamefont {Ni}\ \emph {et~al.}(2010)\citenamefont {Ni},
  \citenamefont {Ospelkaus}, \citenamefont {Wang}, \citenamefont
  {Qu{\'{e}}m{\'{e}}ner}, \citenamefont {Neyenhuis}, \citenamefont
  {de~Miranda}, \citenamefont {Bohn}, \citenamefont {Ye},\ and\ \citenamefont
  {Jin}}]{Ni2010}%
  \BibitemOpen
  \bibfield  {author} {\bibinfo {author} {\bibfnamefont {K.-K.}\ \bibnamefont
  {Ni}}, \bibinfo {author} {\bibfnamefont {S.}~\bibnamefont {Ospelkaus}},
  \bibinfo {author} {\bibfnamefont {D.}~\bibnamefont {Wang}}, \bibinfo {author}
  {\bibfnamefont {G.}~\bibnamefont {Qu{\'{e}}m{\'{e}}ner}}, \bibinfo {author}
  {\bibfnamefont {B.}~\bibnamefont {Neyenhuis}}, \bibinfo {author}
  {\bibfnamefont {M.~H.~G.}\ \bibnamefont {de~Miranda}}, \bibinfo {author}
  {\bibfnamefont {J.~L.}\ \bibnamefont {Bohn}}, \bibinfo {author}
  {\bibfnamefont {J.}~\bibnamefont {Ye}}, \ and\ \bibinfo {author}
  {\bibfnamefont {D.~S.}\ \bibnamefont {Jin}},\ }\href {\doibase
  10.1038/nature08953} {\bibfield  {journal} {\bibinfo  {journal} {Nature}\
  }\textbf {\bibinfo {volume} {464}},\ \bibinfo {pages} {1324} (\bibinfo {year}
  {2010})}\BibitemShut {NoStop}%
\bibitem [{\citenamefont {de~Miranda}\ \emph {et~al.}(2011)\citenamefont
  {de~Miranda}, \citenamefont {Chotia}, \citenamefont {Neyenhuis},
  \citenamefont {D.}, \citenamefont {Qu{\'{e}}m{\'{e}}ner}, \citenamefont
  {Ospelkaus}, \citenamefont {Bohn}, \citenamefont {Ye},\ and\ \citenamefont
  {Jin}}]{DeMiranda2011NP}%
  \BibitemOpen
  \bibfield  {author} {\bibinfo {author} {\bibfnamefont {M.~H.~G.}\
  \bibnamefont {de~Miranda}}, \bibinfo {author} {\bibfnamefont
  {A.}~\bibnamefont {Chotia}}, \bibinfo {author} {\bibfnamefont
  {B.}~\bibnamefont {Neyenhuis}}, \bibinfo {author} {\bibfnamefont
  {W.}~\bibnamefont {D.}}, \bibinfo {author} {\bibfnamefont {G.}~\bibnamefont
  {Qu{\'{e}}m{\'{e}}ner}}, \bibinfo {author} {\bibfnamefont {S.}~\bibnamefont
  {Ospelkaus}}, \bibinfo {author} {\bibfnamefont {J.~L.}\ \bibnamefont {Bohn}},
  \bibinfo {author} {\bibfnamefont {J.}~\bibnamefont {Ye}}, \ and\ \bibinfo
  {author} {\bibfnamefont {D.~S.}\ \bibnamefont {Jin}},\ }\href {\doibase
  10.1038/nphys1939} {\bibfield  {journal} {\bibinfo  {journal} {Nat. Phys.}\
  }\textbf {\bibinfo {volume} {7}},\ \bibinfo {pages} {502} (\bibinfo {year}
  {2011})}\BibitemShut {NoStop}%
\bibitem [{\citenamefont {Macia}\ \emph {et~al.}(2014)\citenamefont {Macia},
  \citenamefont {Astrakharchik}, \citenamefont {Mazzanti}, \citenamefont
  {Giorgini},\ and\ \citenamefont {Boronat}}]{Macia2014PRA_1}%
  \BibitemOpen
  \bibfield  {author} {\bibinfo {author} {\bibfnamefont {A.}~\bibnamefont
  {Macia}}, \bibinfo {author} {\bibfnamefont {G.~E.}\ \bibnamefont
  {Astrakharchik}}, \bibinfo {author} {\bibfnamefont {F.}~\bibnamefont
  {Mazzanti}}, \bibinfo {author} {\bibfnamefont {S.}~\bibnamefont {Giorgini}},
  \ and\ \bibinfo {author} {\bibfnamefont {J.}~\bibnamefont {Boronat}},\ }\href
  {\doibase 10.1103/PhysRevA.90.043623} {\bibfield  {journal} {\bibinfo
  {journal} {Phys. Rev. A}\ }\textbf {\bibinfo {volume} {90}},\ \bibinfo
  {pages} {043623} (\bibinfo {year} {2014})}\BibitemShut {NoStop}%
\bibitem [{\citenamefont {Bombin}\ \emph {et~al.}(2017)\citenamefont {Bombin},
  \citenamefont {Boronat},\ and\ \citenamefont {Mazzanti}}]{Bombin2017PRL}%
  \BibitemOpen
  \bibfield  {author} {\bibinfo {author} {\bibfnamefont {R.}~\bibnamefont
  {Bombin}}, \bibinfo {author} {\bibfnamefont {J.}~\bibnamefont {Boronat}}, \
  and\ \bibinfo {author} {\bibfnamefont {F.}~\bibnamefont {Mazzanti}},\ }\href
  {\doibase 10.1103/PhysRevLett.119.250402} {\bibfield  {journal} {\bibinfo
  {journal} {Phys. Rev. Lett.}\ }\textbf {\bibinfo {volume} {119}},\ \bibinfo
  {pages} {250402} (\bibinfo {year} {2017})}\BibitemShut {NoStop}%
\bibitem [{\citenamefont {Bruun}\ and\ \citenamefont
  {Taylor}(2008)}]{Bruun2008PRL}%
  \BibitemOpen
  \bibfield  {author} {\bibinfo {author} {\bibfnamefont {G.~M.}\ \bibnamefont
  {Bruun}}\ and\ \bibinfo {author} {\bibfnamefont {E.}~\bibnamefont {Taylor}},\
  }\href {\doibase 10.1103/PhysRevLett.101.245301} {\bibfield  {journal}
  {\bibinfo  {journal} {Phys. Rev. Lett.}\ }\textbf {\bibinfo {volume} {101}},\
  \bibinfo {pages} {245301} (\bibinfo {year} {2008})}\BibitemShut {NoStop}%
\bibitem [{\citenamefont {Cooper}\ and\ \citenamefont
  {Shlyapnikov}(2009)}]{Cooper2009PRL}%
  \BibitemOpen
  \bibfield  {author} {\bibinfo {author} {\bibfnamefont {N.~R.}\ \bibnamefont
  {Cooper}}\ and\ \bibinfo {author} {\bibfnamefont {G.~V.}\ \bibnamefont
  {Shlyapnikov}},\ }\href {\doibase 10.1103/PhysRevLett.103.155302} {\bibfield
  {journal} {\bibinfo  {journal} {Phys. Rev. Lett.}\ }\textbf {\bibinfo
  {volume} {103}},\ \bibinfo {pages} {155302} (\bibinfo {year}
  {2009})}\BibitemShut {NoStop}%
\bibitem [{\citenamefont {B\"uchler}\ \emph {et~al.}(2007)\citenamefont
  {B\"uchler}, \citenamefont {Demler}, \citenamefont {Lukin}, \citenamefont
  {Micheli}, \citenamefont {Prokof'ev}, \citenamefont {Pupillo},\ and\
  \citenamefont {Zoller}}]{Buchler2007PRL}%
  \BibitemOpen
  \bibfield  {author} {\bibinfo {author} {\bibfnamefont {H.~P.}\ \bibnamefont
  {B\"uchler}}, \bibinfo {author} {\bibfnamefont {E.}~\bibnamefont {Demler}},
  \bibinfo {author} {\bibfnamefont {M.}~\bibnamefont {Lukin}}, \bibinfo
  {author} {\bibfnamefont {A.}~\bibnamefont {Micheli}}, \bibinfo {author}
  {\bibfnamefont {N.}~\bibnamefont {Prokof'ev}}, \bibinfo {author}
  {\bibfnamefont {G.}~\bibnamefont {Pupillo}}, \ and\ \bibinfo {author}
  {\bibfnamefont {P.}~\bibnamefont {Zoller}},\ }\href {\doibase
  10.1103/PhysRevLett.98.060404} {\bibfield  {journal} {\bibinfo  {journal}
  {Phys. Rev. Lett.}\ }\textbf {\bibinfo {volume} {98}},\ \bibinfo {pages}
  {060404} (\bibinfo {year} {2007})}\BibitemShut {NoStop}%
\bibitem [{\citenamefont {Astrakharchik}\ \emph
  {et~al.}(2007{\natexlab{a}})\citenamefont {Astrakharchik}, \citenamefont
  {Boronat}, \citenamefont {Kurbakov},\ and\ \citenamefont
  {Lozovik}}]{Astrakharchik2007PRL}%
  \BibitemOpen
  \bibfield  {author} {\bibinfo {author} {\bibfnamefont {G.~E.}\ \bibnamefont
  {Astrakharchik}}, \bibinfo {author} {\bibfnamefont {J.}~\bibnamefont
  {Boronat}}, \bibinfo {author} {\bibfnamefont {I.~L.}\ \bibnamefont
  {Kurbakov}}, \ and\ \bibinfo {author} {\bibfnamefont {Y.~E.}\ \bibnamefont
  {Lozovik}},\ }\href {\doibase 10.1103/PhysRevLett.98.060405} {\bibfield
  {journal} {\bibinfo  {journal} {Phys. Rev. Lett.}\ }\textbf {\bibinfo
  {volume} {98}},\ \bibinfo {pages} {060405} (\bibinfo {year}
  {2007}{\natexlab{a}})}\BibitemShut {NoStop}%
\bibitem [{\citenamefont {Matveeva}\ and\ \citenamefont
  {Giorgini}(2012)}]{Matveeva2012PRL}%
  \BibitemOpen
  \bibfield  {author} {\bibinfo {author} {\bibfnamefont {N.}~\bibnamefont
  {Matveeva}}\ and\ \bibinfo {author} {\bibfnamefont {S.}~\bibnamefont
  {Giorgini}},\ }\href {\doibase 10.1103/PhysRevLett.109.200401} {\bibfield
  {journal} {\bibinfo  {journal} {Phys. Rev. Lett.}\ }\textbf {\bibinfo
  {volume} {109}},\ \bibinfo {pages} {200401} (\bibinfo {year}
  {2012})}\BibitemShut {NoStop}%
\bibitem [{\citenamefont {Hammond}\ \emph {et~al.}(1994)\citenamefont
  {Hammond}, \citenamefont {Lester},\ and\ \citenamefont
  {Reynolds}}]{Hammond1994}%
  \BibitemOpen
  \bibfield  {author} {\bibinfo {author} {\bibfnamefont {B.~L.}\ \bibnamefont
  {Hammond}}, \bibinfo {author} {\bibfnamefont {W.~A.}\ \bibnamefont {Lester}},
  \ and\ \bibinfo {author} {\bibfnamefont {P.~J.}\ \bibnamefont {Reynolds}},\
  }\href {\doibase 10.1142/1170} {\emph {\bibinfo {title} {{Monte Carlo Methods
  in Ab Initio Quantum Chemistry}}}}\ (\bibinfo  {publisher} {World
  Scientific},\ \bibinfo {year} {1994})\BibitemShut {NoStop}%
\bibitem [{\citenamefont {Foulkes}\ \emph {et~al.}(2001)\citenamefont
  {Foulkes}, \citenamefont {Mitas}, \citenamefont {Needs},\ and\ \citenamefont
  {Rajagopal}}]{Foulkes2001RMP}%
  \BibitemOpen
  \bibfield  {author} {\bibinfo {author} {\bibfnamefont {W.~M.~C.}\
  \bibnamefont {Foulkes}}, \bibinfo {author} {\bibfnamefont {L.}~\bibnamefont
  {Mitas}}, \bibinfo {author} {\bibfnamefont {R.~J.}\ \bibnamefont {Needs}}, \
  and\ \bibinfo {author} {\bibfnamefont {G.}~\bibnamefont {Rajagopal}},\ }\href
  {\doibase 10.1103/RevModPhys.73.33} {\bibfield  {journal} {\bibinfo
  {journal} {Rev. Mod. Phys.}\ }\textbf {\bibinfo {volume} {73}},\ \bibinfo
  {pages} {33} (\bibinfo {year} {2001})}\BibitemShut {NoStop}%
\bibitem [{\citenamefont {Koloren\v{c}}\ and\ \citenamefont
  {Mitas}(2011)}]{Kolorenc2011RPP}%
  \BibitemOpen
  \bibfield  {author} {\bibinfo {author} {\bibfnamefont {J.}~\bibnamefont
  {Koloren\v{c}}}\ and\ \bibinfo {author} {\bibfnamefont {L.}~\bibnamefont
  {Mitas}},\ }\href {\doibase 10.1088/0034-4885/74/2/026502} {\bibfield
  {journal} {\bibinfo  {journal} {Rep. Progr. Phys.}\ }\textbf {\bibinfo
  {volume} {74}},\ \bibinfo {pages} {026502} (\bibinfo {year}
  {2011})}\BibitemShut {NoStop}%
\bibitem [{\citenamefont {Bloch}(1929)}]{Bloch1929ZP}%
  \BibitemOpen
  \bibfield  {author} {\bibinfo {author} {\bibfnamefont {F.}~\bibnamefont
  {Bloch}},\ }\href {\doibase 10.1007/BF01340281} {\bibfield  {journal}
  {\bibinfo  {journal} {Z. Phys.}\ }\textbf {\bibinfo {volume} {57}},\ \bibinfo
  {pages} {545} (\bibinfo {year} {1929})}\BibitemShut {NoStop}%
\bibitem [{\citenamefont {Ceperley}(1978)}]{Ceperley1978PRB}%
  \BibitemOpen
  \bibfield  {author} {\bibinfo {author} {\bibfnamefont {D.}~\bibnamefont
  {Ceperley}},\ }\href {\doibase 10.1103/PhysRevB.18.3126} {\bibfield
  {journal} {\bibinfo  {journal} {Phys. Rev. B}\ }\textbf {\bibinfo {volume}
  {18}},\ \bibinfo {pages} {3126} (\bibinfo {year} {1978})}\BibitemShut
  {NoStop}%
\bibitem [{\citenamefont {Ceperley}\ and\ \citenamefont
  {Alder}(1980)}]{Ceperley1980PRL}%
  \BibitemOpen
  \bibfield  {author} {\bibinfo {author} {\bibfnamefont {D.~M.}\ \bibnamefont
  {Ceperley}}\ and\ \bibinfo {author} {\bibfnamefont {B.~J.}\ \bibnamefont
  {Alder}},\ }\href {\doibase 10.1103/PhysRevLett.45.566} {\bibfield  {journal}
  {\bibinfo  {journal} {Phys. Rev. Lett.}\ }\textbf {\bibinfo {volume} {45}},\
  \bibinfo {pages} {566} (\bibinfo {year} {1980})}\BibitemShut {NoStop}%
\bibitem [{\citenamefont {Alder}\ \emph {et~al.}(1982)\citenamefont {Alder},
  \citenamefont {Ceperley},\ and\ \citenamefont {Pollock}}]{Alder1982IJQC}%
  \BibitemOpen
  \bibfield  {author} {\bibinfo {author} {\bibfnamefont {B.~J.}\ \bibnamefont
  {Alder}}, \bibinfo {author} {\bibfnamefont {D.~M.}\ \bibnamefont {Ceperley}},
  \ and\ \bibinfo {author} {\bibfnamefont {E.~L.}\ \bibnamefont {Pollock}},\
  }\href {\doibase 10.1002/qua.560220808} {\bibfield  {journal} {\bibinfo
  {journal} {Int. J. Quantum Chem.}\ }\textbf {\bibinfo {volume} {22}},\
  \bibinfo {pages} {49} (\bibinfo {year} {1982})}\BibitemShut {NoStop}%
\bibitem [{\citenamefont {Tanatar}\ and\ \citenamefont
  {Ceperley}(1989)}]{Tanatar1989PRB}%
  \BibitemOpen
  \bibfield  {author} {\bibinfo {author} {\bibfnamefont {B.}~\bibnamefont
  {Tanatar}}\ and\ \bibinfo {author} {\bibfnamefont {D.~M.}\ \bibnamefont
  {Ceperley}},\ }\href {\doibase 10.1103/PhysRevB.39.5005} {\bibfield
  {journal} {\bibinfo  {journal} {Phys. Rev. B}\ }\textbf {\bibinfo {volume}
  {39}},\ \bibinfo {pages} {5005} (\bibinfo {year} {1989})}\BibitemShut
  {NoStop}%
\bibitem [{\citenamefont {Ortiz}\ \emph {et~al.}(1999)\citenamefont {Ortiz},
  \citenamefont {Harris},\ and\ \citenamefont {Ballone}}]{Ortiz1999PRL}%
  \BibitemOpen
  \bibfield  {author} {\bibinfo {author} {\bibfnamefont {G.}~\bibnamefont
  {Ortiz}}, \bibinfo {author} {\bibfnamefont {M.}~\bibnamefont {Harris}}, \
  and\ \bibinfo {author} {\bibfnamefont {P.}~\bibnamefont {Ballone}},\ }\href
  {\doibase 10.1103/PhysRevLett.82.5317} {\bibfield  {journal} {\bibinfo
  {journal} {Phys. Rev. Lett.}\ }\textbf {\bibinfo {volume} {82}},\ \bibinfo
  {pages} {5317} (\bibinfo {year} {1999})}\BibitemShut {NoStop}%
\bibitem [{\citenamefont {Zong}\ \emph {et~al.}(2002)\citenamefont {Zong},
  \citenamefont {Lin},\ and\ \citenamefont {Ceperley}}]{Zong2002PRE}%
  \BibitemOpen
  \bibfield  {author} {\bibinfo {author} {\bibfnamefont {F.~H.}\ \bibnamefont
  {Zong}}, \bibinfo {author} {\bibfnamefont {C.}~\bibnamefont {Lin}}, \ and\
  \bibinfo {author} {\bibfnamefont {D.~M.}\ \bibnamefont {Ceperley}},\ }\href
  {\doibase 10.1103/PhysRevE.66.036703} {\bibfield  {journal} {\bibinfo
  {journal} {Phys. Rev. E}\ }\textbf {\bibinfo {volume} {66}},\ \bibinfo
  {pages} {036703} (\bibinfo {year} {2002})}\BibitemShut {NoStop}%
\bibitem [{\citenamefont {Drummond}\ and\ \citenamefont
  {Needs}(2009)}]{Drummond2009PRL}%
  \BibitemOpen
  \bibfield  {author} {\bibinfo {author} {\bibfnamefont {N.~D.}\ \bibnamefont
  {Drummond}}\ and\ \bibinfo {author} {\bibfnamefont {R.~J.}\ \bibnamefont
  {Needs}},\ }\href {\doibase 10.1103/PhysRevLett.102.126402} {\bibfield
  {journal} {\bibinfo  {journal} {Phys. Rev. Lett.}\ }\textbf {\bibinfo
  {volume} {102}},\ \bibinfo {pages} {126402} (\bibinfo {year}
  {2009})}\BibitemShut {NoStop}%
\bibitem [{\citenamefont {Manousakis}\ \emph {et~al.}(1983)\citenamefont
  {Manousakis}, \citenamefont {Fantoni}, \citenamefont {Pandharipande},\ and\
  \citenamefont {Usmani}}]{Manousakis1983PRB}%
  \BibitemOpen
  \bibfield  {author} {\bibinfo {author} {\bibfnamefont {E.}~\bibnamefont
  {Manousakis}}, \bibinfo {author} {\bibfnamefont {S.}~\bibnamefont {Fantoni}},
  \bibinfo {author} {\bibfnamefont {V.~R.}\ \bibnamefont {Pandharipande}}, \
  and\ \bibinfo {author} {\bibfnamefont {Q.~N.}\ \bibnamefont {Usmani}},\
  }\href {\doibase 10.1103/PhysRevB.28.3770} {\bibfield  {journal} {\bibinfo
  {journal} {Phys. Rev. B}\ }\textbf {\bibinfo {volume} {28}},\ \bibinfo
  {pages} {3770} (\bibinfo {year} {1983})}\BibitemShut {NoStop}%
\bibitem [{\citenamefont {Zong}\ \emph {et~al.}(2003)\citenamefont {Zong},
  \citenamefont {Ceperley}, \citenamefont {Moroni},\ and\ \citenamefont
  {Fantoni}}]{Zong2003MP}%
  \BibitemOpen
  \bibfield  {author} {\bibinfo {author} {\bibfnamefont {F.~H.}\ \bibnamefont
  {Zong}}, \bibinfo {author} {\bibfnamefont {D.~M.}\ \bibnamefont {Ceperley}},
  \bibinfo {author} {\bibfnamefont {S.}~\bibnamefont {Moroni}}, \ and\ \bibinfo
  {author} {\bibfnamefont {S.}~\bibnamefont {Fantoni}},\ }\href {\doibase
  10.1080/0026897031000085119} {\bibfield  {journal} {\bibinfo  {journal} {Mol.
  Phys.}\ }\textbf {\bibinfo {volume} {101}},\ \bibinfo {pages} {1705}
  (\bibinfo {year} {2003})}\BibitemShut {NoStop}%
\bibitem [{\citenamefont {Nava}\ \emph {et~al.}(2012)\citenamefont {Nava},
  \citenamefont {Motta}, \citenamefont {Galli}, \citenamefont {Vitali},\ and\
  \citenamefont {Moroni}}]{Nava2012PRB}%
  \BibitemOpen
  \bibfield  {author} {\bibinfo {author} {\bibfnamefont {M.}~\bibnamefont
  {Nava}}, \bibinfo {author} {\bibfnamefont {A.}~\bibnamefont {Motta}},
  \bibinfo {author} {\bibfnamefont {D.~E.}\ \bibnamefont {Galli}}, \bibinfo
  {author} {\bibfnamefont {E.}~\bibnamefont {Vitali}}, \ and\ \bibinfo {author}
  {\bibfnamefont {S.}~\bibnamefont {Moroni}},\ }\href {\doibase
  10.1103/PhysRevB.85.184401} {\bibfield  {journal} {\bibinfo  {journal} {Phys.
  Rev. B}\ }\textbf {\bibinfo {volume} {85}},\ \bibinfo {pages} {184401}
  (\bibinfo {year} {2012})}\BibitemShut {NoStop}%
\bibitem [{\citenamefont {Pilati}\ \emph {et~al.}(2010)\citenamefont {Pilati},
  \citenamefont {Bertaina}, \citenamefont {Giorgini},\ and\ \citenamefont
  {Troyer}}]{Pilati2010PRL}%
  \BibitemOpen
  \bibfield  {author} {\bibinfo {author} {\bibfnamefont {S.}~\bibnamefont
  {Pilati}}, \bibinfo {author} {\bibfnamefont {G.}~\bibnamefont {Bertaina}},
  \bibinfo {author} {\bibfnamefont {S.}~\bibnamefont {Giorgini}}, \ and\
  \bibinfo {author} {\bibfnamefont {M.}~\bibnamefont {Troyer}},\ }\href
  {\doibase 10.1103/PhysRevLett.105.030405} {\bibfield  {journal} {\bibinfo
  {journal} {Phys. Rev. Lett.}\ }\textbf {\bibinfo {volume} {105}},\ \bibinfo
  {pages} {030405} (\bibinfo {year} {2010})}\BibitemShut {NoStop}%
\bibitem [{\citenamefont {Chang}\ \emph {et~al.}(2011)\citenamefont {Chang},
  \citenamefont {Randeria},\ and\ \citenamefont {Trivedi}}]{Chang2011PNAS}%
  \BibitemOpen
  \bibfield  {author} {\bibinfo {author} {\bibfnamefont {S.-Y.}\ \bibnamefont
  {Chang}}, \bibinfo {author} {\bibfnamefont {M.}~\bibnamefont {Randeria}}, \
  and\ \bibinfo {author} {\bibfnamefont {N.}~\bibnamefont {Trivedi}},\ }\href
  {\doibase 10.1073/pnas.1011990108} {\bibfield  {journal} {\bibinfo  {journal}
  {Proc. Nat. Acad. Sci.}\ }\textbf {\bibinfo {volume} {108}},\ \bibinfo
  {pages} {51} (\bibinfo {year} {2011})}\BibitemShut {NoStop}%
\bibitem [{\citenamefont {Arias~de Saavedra}\ \emph {et~al.}(2012)\citenamefont
  {Arias~de Saavedra}, \citenamefont {Mazzanti}, \citenamefont {Boronat},\ and\
  \citenamefont {Polls}}]{AriasDeSaavedra2012PRA}%
  \BibitemOpen
  \bibfield  {author} {\bibinfo {author} {\bibfnamefont {F.}~\bibnamefont
  {Arias~de Saavedra}}, \bibinfo {author} {\bibfnamefont {F.}~\bibnamefont
  {Mazzanti}}, \bibinfo {author} {\bibfnamefont {J.}~\bibnamefont {Boronat}}, \
  and\ \bibinfo {author} {\bibfnamefont {A.}~\bibnamefont {Polls}},\ }\href
  {\doibase 10.1103/PhysRevA.85.033615} {\bibfield  {journal} {\bibinfo
  {journal} {Phys. Rev. A}\ }\textbf {\bibinfo {volume} {85}},\ \bibinfo
  {pages} {033615} (\bibinfo {year} {2012})}\BibitemShut {NoStop}%
\bibitem [{\citenamefont {Pilati}\ \emph {et~al.}(2014)\citenamefont {Pilati},
  \citenamefont {Zintchenko},\ and\ \citenamefont {Troyer}}]{Pilati2014PRL}%
  \BibitemOpen
  \bibfield  {author} {\bibinfo {author} {\bibfnamefont {S.}~\bibnamefont
  {Pilati}}, \bibinfo {author} {\bibfnamefont {I.}~\bibnamefont {Zintchenko}},
  \ and\ \bibinfo {author} {\bibfnamefont {M.}~\bibnamefont {Troyer}},\ }\href
  {\doibase 10.1103/PhysRevLett.112.015301} {\bibfield  {journal} {\bibinfo
  {journal} {Phys. Rev. Lett.}\ }\textbf {\bibinfo {volume} {112}},\ \bibinfo
  {pages} {015301} (\bibinfo {year} {2014})}\BibitemShut {NoStop}%
\bibitem [{\citenamefont {He}\ \emph {et~al.}(2016)\citenamefont {He},
  \citenamefont {Liu}, \citenamefont {Huang},\ and\ \citenamefont
  {Hu}}]{He2016PRA}%
  \BibitemOpen
  \bibfield  {author} {\bibinfo {author} {\bibfnamefont {L.}~\bibnamefont
  {He}}, \bibinfo {author} {\bibfnamefont {X.-J.}\ \bibnamefont {Liu}},
  \bibinfo {author} {\bibfnamefont {X.-G.}\ \bibnamefont {Huang}}, \ and\
  \bibinfo {author} {\bibfnamefont {H.}~\bibnamefont {Hu}},\ }\href {\doibase
  10.1103/PhysRevA.93.063629} {\bibfield  {journal} {\bibinfo  {journal} {Phys.
  Rev. A}\ }\textbf {\bibinfo {volume} {93}},\ \bibinfo {pages} {063629}
  (\bibinfo {year} {2016})}\BibitemShut {NoStop}%
\bibitem [{\citenamefont {Vermeyen}\ \emph {et~al.}(2018)\citenamefont
  {Vermeyen}, \citenamefont {S\'a~de Melo},\ and\ \citenamefont
  {Tempere}}]{Vermeyen2018PRA}%
  \BibitemOpen
  \bibfield  {author} {\bibinfo {author} {\bibfnamefont {E.}~\bibnamefont
  {Vermeyen}}, \bibinfo {author} {\bibfnamefont {C.~A.~R.}\ \bibnamefont
  {S\'a~de Melo}}, \ and\ \bibinfo {author} {\bibfnamefont {J.}~\bibnamefont
  {Tempere}},\ }\href {\doibase 10.1103/PhysRevA.98.023635} {\bibfield
  {journal} {\bibinfo  {journal} {Phys. Rev. A}\ }\textbf {\bibinfo {volume}
  {98}},\ \bibinfo {pages} {023635} (\bibinfo {year} {2018})}\BibitemShut
  {NoStop}%
\bibitem [{\citenamefont {Zintchenko}\ \emph {et~al.}(2016)\citenamefont
  {Zintchenko}, \citenamefont {Wang},\ and\ \citenamefont
  {Troyer}}]{Zintchenko2016EPJB}%
  \BibitemOpen
  \bibfield  {author} {\bibinfo {author} {\bibfnamefont {I.}~\bibnamefont
  {Zintchenko}}, \bibinfo {author} {\bibfnamefont {L.}~\bibnamefont {Wang}}, \
  and\ \bibinfo {author} {\bibfnamefont {M.}~\bibnamefont {Troyer}},\ }\href
  {\doibase 10.1140/epjb/e2016-70302-5} {\bibfield  {journal} {\bibinfo
  {journal} {Eur. Phys. J. B}\ }\textbf {\bibinfo {volume} {89}},\ \bibinfo
  {pages} {180} (\bibinfo {year} {2016})}\BibitemShut {NoStop}%
\bibitem [{\citenamefont {Stoner}(1938)}]{Stoner1938PRSL}%
  \BibitemOpen
  \bibfield  {author} {\bibinfo {author} {\bibfnamefont {E.~C.}\ \bibnamefont
  {Stoner}},\ }\href {\doibase 10.1098/rspa.1938.0066} {\bibfield  {journal}
  {\bibinfo  {journal} {Proc. R. Soc. London, Ser. A}\ }\textbf {\bibinfo
  {volume} {165}},\ \bibinfo {pages} {372} (\bibinfo {year}
  {1938})}\BibitemShut {NoStop}%
\bibitem [{\citenamefont {Valtolina}\ \emph {et~al.}(2017)\citenamefont
  {Valtolina}, \citenamefont {Scazza}, \citenamefont {Amico}, \citenamefont
  {Burchianti}, \citenamefont {Recati}, \citenamefont {Enss}, \citenamefont
  {Inguscio}, \citenamefont {Zaccanti},\ and\ \citenamefont
  {Roati}}]{Valtolina2017NP}%
  \BibitemOpen
  \bibfield  {author} {\bibinfo {author} {\bibfnamefont {G.}~\bibnamefont
  {Valtolina}}, \bibinfo {author} {\bibfnamefont {F.}~\bibnamefont {Scazza}},
  \bibinfo {author} {\bibfnamefont {A.}~\bibnamefont {Amico}}, \bibinfo
  {author} {\bibfnamefont {A.}~\bibnamefont {Burchianti}}, \bibinfo {author}
  {\bibfnamefont {A.}~\bibnamefont {Recati}}, \bibinfo {author} {\bibfnamefont
  {T.}~\bibnamefont {Enss}}, \bibinfo {author} {\bibfnamefont {M.}~\bibnamefont
  {Inguscio}}, \bibinfo {author} {\bibfnamefont {M.}~\bibnamefont {Zaccanti}},
  \ and\ \bibinfo {author} {\bibfnamefont {G.}~\bibnamefont {Roati}},\ }\href
  {\doibase 10.1038/nphys4108} {\bibfield  {journal} {\bibinfo  {journal} {Nat.
  Phys.}\ }\textbf {\bibinfo {volume} {13}},\ \bibinfo {pages} {704} (\bibinfo
  {year} {2017})}\BibitemShut {NoStop}%
\bibitem [{\citenamefont {Jo}\ \emph {et~al.}(2009)\citenamefont {Jo},
  \citenamefont {Lee}, \citenamefont {Choi}, \citenamefont {Christensen},
  \citenamefont {Kim}, \citenamefont {Thywissen}, \citenamefont {Pritchard},\
  and\ \citenamefont {Ketterle}}]{Jo2009}%
  \BibitemOpen
  \bibfield  {author} {\bibinfo {author} {\bibfnamefont {G.-B.}\ \bibnamefont
  {Jo}}, \bibinfo {author} {\bibfnamefont {Y.-R.}\ \bibnamefont {Lee}},
  \bibinfo {author} {\bibfnamefont {J.-H.}\ \bibnamefont {Choi}}, \bibinfo
  {author} {\bibfnamefont {C.~A.}\ \bibnamefont {Christensen}}, \bibinfo
  {author} {\bibfnamefont {T.~H.}\ \bibnamefont {Kim}}, \bibinfo {author}
  {\bibfnamefont {J.~H.}\ \bibnamefont {Thywissen}}, \bibinfo {author}
  {\bibfnamefont {D.~E.}\ \bibnamefont {Pritchard}}, \ and\ \bibinfo {author}
  {\bibfnamefont {W.}~\bibnamefont {Ketterle}},\ }\href {\doibase
  10.1126/science.1177112} {\bibfield  {journal} {\bibinfo  {journal}
  {Science}\ }\textbf {\bibinfo {volume} {325}},\ \bibinfo {pages} {1521}
  (\bibinfo {year} {2009})}\BibitemShut {NoStop}%
\bibitem [{\citenamefont {Pekker}\ \emph {et~al.}(2011)\citenamefont {Pekker},
  \citenamefont {Babadi}, \citenamefont {Sensarma}, \citenamefont {Zinner},
  \citenamefont {Pollet}, \citenamefont {Zwierlein},\ and\ \citenamefont
  {Demler}}]{Pekker2011PRL}%
  \BibitemOpen
  \bibfield  {author} {\bibinfo {author} {\bibfnamefont {D.}~\bibnamefont
  {Pekker}}, \bibinfo {author} {\bibfnamefont {M.}~\bibnamefont {Babadi}},
  \bibinfo {author} {\bibfnamefont {R.}~\bibnamefont {Sensarma}}, \bibinfo
  {author} {\bibfnamefont {N.}~\bibnamefont {Zinner}}, \bibinfo {author}
  {\bibfnamefont {L.}~\bibnamefont {Pollet}}, \bibinfo {author} {\bibfnamefont
  {M.~W.}\ \bibnamefont {Zwierlein}}, \ and\ \bibinfo {author} {\bibfnamefont
  {E.}~\bibnamefont {Demler}},\ }\href {\doibase
  10.1103/PhysRevLett.106.050402} {\bibfield  {journal} {\bibinfo  {journal}
  {Phys. Rev. Lett.}\ }\textbf {\bibinfo {volume} {106}},\ \bibinfo {pages}
  {050402} (\bibinfo {year} {2011})}\BibitemShut {NoStop}%
\bibitem [{\citenamefont {Sanner}\ \emph {et~al.}(2012)\citenamefont {Sanner},
  \citenamefont {Su}, \citenamefont {Huang}, \citenamefont {Keshet},
  \citenamefont {Gillen},\ and\ \citenamefont {Ketterle}}]{Sanner2012PRL}%
  \BibitemOpen
  \bibfield  {author} {\bibinfo {author} {\bibfnamefont {C.}~\bibnamefont
  {Sanner}}, \bibinfo {author} {\bibfnamefont {E.~J.}\ \bibnamefont {Su}},
  \bibinfo {author} {\bibfnamefont {W.}~\bibnamefont {Huang}}, \bibinfo
  {author} {\bibfnamefont {A.}~\bibnamefont {Keshet}}, \bibinfo {author}
  {\bibfnamefont {J.}~\bibnamefont {Gillen}}, \ and\ \bibinfo {author}
  {\bibfnamefont {W.}~\bibnamefont {Ketterle}},\ }\href {\doibase
  10.1103/PhysRevLett.108.240404} {\bibfield  {journal} {\bibinfo  {journal}
  {Phys. Rev. Lett.}\ }\textbf {\bibinfo {volume} {108}},\ \bibinfo {pages}
  {240404} (\bibinfo {year} {2012})}\BibitemShut {NoStop}%
\bibitem [{\citenamefont {Amico}\ \emph {et~al.}(2018)\citenamefont {Amico},
  \citenamefont {Scazza}, \citenamefont {Valtolina}, \citenamefont {Tavares},
  \citenamefont {Ketterle}, \citenamefont {Inguscio}, \citenamefont {Roati},\
  and\ \citenamefont {Zaccanti}}]{Amico2018PRL}%
  \BibitemOpen
  \bibfield  {author} {\bibinfo {author} {\bibfnamefont {A.}~\bibnamefont
  {Amico}}, \bibinfo {author} {\bibfnamefont {F.}~\bibnamefont {Scazza}},
  \bibinfo {author} {\bibfnamefont {G.}~\bibnamefont {Valtolina}}, \bibinfo
  {author} {\bibfnamefont {P.~E.~S.}\ \bibnamefont {Tavares}}, \bibinfo
  {author} {\bibfnamefont {W.}~\bibnamefont {Ketterle}}, \bibinfo {author}
  {\bibfnamefont {M.}~\bibnamefont {Inguscio}}, \bibinfo {author}
  {\bibfnamefont {G.}~\bibnamefont {Roati}}, \ and\ \bibinfo {author}
  {\bibfnamefont {M.}~\bibnamefont {Zaccanti}},\ }\href {\doibase
  10.1103/PhysRevLett.121.253602} {\bibfield  {journal} {\bibinfo  {journal}
  {Phys. Rev. Lett.}\ }\textbf {\bibinfo {volume} {121}},\ \bibinfo {pages}
  {253602} (\bibinfo {year} {2018})}\BibitemShut {NoStop}%
\bibitem [{\citenamefont {Taddei}\ \emph {et~al.}(2015)\citenamefont {Taddei},
  \citenamefont {Ruggeri}, \citenamefont {Moroni},\ and\ \citenamefont
  {Holzmann}}]{Taddei2015PRB}%
  \BibitemOpen
  \bibfield  {author} {\bibinfo {author} {\bibfnamefont {M.}~\bibnamefont
  {Taddei}}, \bibinfo {author} {\bibfnamefont {M.}~\bibnamefont {Ruggeri}},
  \bibinfo {author} {\bibfnamefont {S.}~\bibnamefont {Moroni}}, \ and\ \bibinfo
  {author} {\bibfnamefont {M.}~\bibnamefont {Holzmann}},\ }\href {\doibase
  10.1103/PhysRevB.91.115106} {\bibfield  {journal} {\bibinfo  {journal} {Phys.
  Rev. B}\ }\textbf {\bibinfo {volume} {91}},\ \bibinfo {pages} {115106}
  (\bibinfo {year} {2015})}\BibitemShut {NoStop}%
\bibitem [{\citenamefont {Ruggeri}\ \emph {et~al.}(2018)\citenamefont
  {Ruggeri}, \citenamefont {Moroni},\ and\ \citenamefont
  {Holzmann}}]{Ruggeri2018PRL}%
  \BibitemOpen
  \bibfield  {author} {\bibinfo {author} {\bibfnamefont {M.}~\bibnamefont
  {Ruggeri}}, \bibinfo {author} {\bibfnamefont {S.}~\bibnamefont {Moroni}}, \
  and\ \bibinfo {author} {\bibfnamefont {M.}~\bibnamefont {Holzmann}},\ }\href
  {\doibase 10.1103/PhysRevLett.120.205302} {\bibfield  {journal} {\bibinfo
  {journal} {Phys. Rev. Lett.}\ }\textbf {\bibinfo {volume} {120}},\ \bibinfo
  {pages} {205302} (\bibinfo {year} {2018})}\BibitemShut {NoStop}%
\bibitem [{\citenamefont {Pitaevskii}\ and\ \citenamefont
  {Stringari}(2016)}]{Pitaevskii2016}%
  \BibitemOpen
  \bibfield  {author} {\bibinfo {author} {\bibfnamefont {L.}~\bibnamefont
  {Pitaevskii}}\ and\ \bibinfo {author} {\bibfnamefont {S.}~\bibnamefont
  {Stringari}},\ }\href {\doibase 10.1093/acprof:oso/9780198758884.001.0001}
  {\emph {\bibinfo {title} {{Bose-Einstein Condensation and Superfluidity}}}}\
  (\bibinfo  {publisher} {Oxford University Press},\ \bibinfo {year}
  {2016})\BibitemShut {NoStop}%
\bibitem [{\citenamefont {Drummond}\ \emph {et~al.}(2011)\citenamefont
  {Drummond}, \citenamefont {Cooper}, \citenamefont {Needs},\ and\
  \citenamefont {Shlyapnikov}}]{Drummond2011PRB}%
  \BibitemOpen
  \bibfield  {author} {\bibinfo {author} {\bibfnamefont {N.~D.}\ \bibnamefont
  {Drummond}}, \bibinfo {author} {\bibfnamefont {N.~R.}\ \bibnamefont
  {Cooper}}, \bibinfo {author} {\bibfnamefont {R.~J.}\ \bibnamefont {Needs}}, \
  and\ \bibinfo {author} {\bibfnamefont {G.~V.}\ \bibnamefont {Shlyapnikov}},\
  }\href {\doibase 10.1103/PhysRevB.83.195429} {\bibfield  {journal} {\bibinfo
  {journal} {Phys. Rev. B}\ }\textbf {\bibinfo {volume} {83}},\ \bibinfo
  {pages} {195429} (\bibinfo {year} {2011})}\BibitemShut {NoStop}%
\bibitem [{\citenamefont {Lin}\ \emph {et~al.}(2001)\citenamefont {Lin},
  \citenamefont {Zong},\ and\ \citenamefont {Ceperley}}]{Lin2001PRE}%
  \BibitemOpen
  \bibfield  {author} {\bibinfo {author} {\bibfnamefont {C.}~\bibnamefont
  {Lin}}, \bibinfo {author} {\bibfnamefont {F.~H.}\ \bibnamefont {Zong}}, \
  and\ \bibinfo {author} {\bibfnamefont {D.~M.}\ \bibnamefont {Ceperley}},\
  }\href {\doibase 10.1103/PhysRevE.64.016702} {\bibfield  {journal} {\bibinfo
  {journal} {Phys. Rev. E}\ }\textbf {\bibinfo {volume} {64}},\ \bibinfo
  {pages} {016702} (\bibinfo {year} {2001})}\BibitemShut {NoStop}%
\bibitem [{\citenamefont {Anderson}(1995)}]{Anderson1995IRPC}%
  \BibitemOpen
  \bibfield  {author} {\bibinfo {author} {\bibfnamefont {J.~B.}\ \bibnamefont
  {Anderson}},\ }\href {\doibase 10.1080/01442359509353305} {\bibfield
  {journal} {\bibinfo  {journal} {Int. Rev. Phys. Chem.}\ }\textbf {\bibinfo
  {volume} {14}},\ \bibinfo {pages} {85} (\bibinfo {year} {1995})}\BibitemShut
  {NoStop}%
\bibitem [{\citenamefont {Macia}\ \emph {et~al.}(2011)\citenamefont {Macia},
  \citenamefont {Mazzanti}, \citenamefont {Boronat},\ and\ \citenamefont
  {Zillich}}]{Macia2011PRA}%
  \BibitemOpen
  \bibfield  {author} {\bibinfo {author} {\bibfnamefont {A.}~\bibnamefont
  {Macia}}, \bibinfo {author} {\bibfnamefont {F.}~\bibnamefont {Mazzanti}},
  \bibinfo {author} {\bibfnamefont {J.}~\bibnamefont {Boronat}}, \ and\
  \bibinfo {author} {\bibfnamefont {R.~E.}\ \bibnamefont {Zillich}},\ }\href
  {\doibase 10.1103/PhysRevA.84.033625} {\bibfield  {journal} {\bibinfo
  {journal} {Phys. Rev. A}\ }\textbf {\bibinfo {volume} {84}},\ \bibinfo
  {pages} {033625} (\bibinfo {year} {2011})}\BibitemShut {NoStop}%
\bibitem [{\citenamefont {Grau}\ \emph {et~al.}(2002)\citenamefont {Grau},
  \citenamefont {Boronat},\ and\ \citenamefont {Casulleras}}]{Grau2002PRL}%
  \BibitemOpen
  \bibfield  {author} {\bibinfo {author} {\bibfnamefont {V.}~\bibnamefont
  {Grau}}, \bibinfo {author} {\bibfnamefont {J.}~\bibnamefont {Boronat}}, \
  and\ \bibinfo {author} {\bibfnamefont {J.}~\bibnamefont {Casulleras}},\
  }\href {\doibase 10.1103/PhysRevLett.89.045301} {\bibfield  {journal}
  {\bibinfo  {journal} {Phys. Rev. Lett.}\ }\textbf {\bibinfo {volume} {89}},\
  \bibinfo {pages} {045301} (\bibinfo {year} {2002})}\BibitemShut {NoStop}%
\bibitem [{\citenamefont {Casulleras}\ and\ \citenamefont
  {Boronat}(2000)}]{Casulleras2000PRL}%
  \BibitemOpen
  \bibfield  {author} {\bibinfo {author} {\bibfnamefont {J.}~\bibnamefont
  {Casulleras}}\ and\ \bibinfo {author} {\bibfnamefont {J.}~\bibnamefont
  {Boronat}},\ }\href {\doibase 10.1103/PhysRevLett.84.3121} {\bibfield
  {journal} {\bibinfo  {journal} {Phys. Rev. Lett.}\ }\textbf {\bibinfo
  {volume} {84}},\ \bibinfo {pages} {3121} (\bibinfo {year}
  {2000})}\BibitemShut {NoStop}%
\bibitem [{\citenamefont {Holzmann}\ \emph {et~al.}(2006)\citenamefont
  {Holzmann}, \citenamefont {Bernu},\ and\ \citenamefont
  {Ceperley}}]{Holzmann2006PRB}%
  \BibitemOpen
  \bibfield  {author} {\bibinfo {author} {\bibfnamefont {M.}~\bibnamefont
  {Holzmann}}, \bibinfo {author} {\bibfnamefont {B.}~\bibnamefont {Bernu}}, \
  and\ \bibinfo {author} {\bibfnamefont {D.~M.}\ \bibnamefont {Ceperley}},\
  }\href {\doibase 10.1103/PhysRevB.74.104510} {\bibfield  {journal} {\bibinfo
  {journal} {Phys. Rev. B}\ }\textbf {\bibinfo {volume} {74}},\ \bibinfo
  {pages} {104510} (\bibinfo {year} {2006})}\BibitemShut {NoStop}%
\bibitem [{\citenamefont {Casulleras}\ and\ \citenamefont
  {Boronat}(1995)}]{Casulleras1995PRB}%
  \BibitemOpen
  \bibfield  {author} {\bibinfo {author} {\bibfnamefont {J.}~\bibnamefont
  {Casulleras}}\ and\ \bibinfo {author} {\bibfnamefont {J.}~\bibnamefont
  {Boronat}},\ }\href {\doibase 10.1103/PhysRevB.52.3654} {\bibfield  {journal}
  {\bibinfo  {journal} {Phys. Rev. B}\ }\textbf {\bibinfo {volume} {52}},\
  \bibinfo {pages} {3654} (\bibinfo {year} {1995})}\BibitemShut {NoStop}%
\bibitem [{\citenamefont {Lu}\ and\ \citenamefont
  {Shlyapnikov}(2012)}]{Lu2012PRA}%
  \BibitemOpen
  \bibfield  {author} {\bibinfo {author} {\bibfnamefont {Z.-K.}\ \bibnamefont
  {Lu}}\ and\ \bibinfo {author} {\bibfnamefont {G.~V.}\ \bibnamefont
  {Shlyapnikov}},\ }\href {\doibase 10.1103/PhysRevA.85.023614} {\bibfield
  {journal} {\bibinfo  {journal} {Phys. Rev. A}\ }\textbf {\bibinfo {volume}
  {85}},\ \bibinfo {pages} {023614} (\bibinfo {year} {2012})}\BibitemShut
  {NoStop}%
\bibitem [{\citenamefont {Lange}\ \emph {et~al.}(2016)\citenamefont {Lange},
  \citenamefont {Krieg},\ and\ \citenamefont {Kopietz}}]{Lange2016PRA}%
  \BibitemOpen
  \bibfield  {author} {\bibinfo {author} {\bibfnamefont {P.}~\bibnamefont
  {Lange}}, \bibinfo {author} {\bibfnamefont {J.}~\bibnamefont {Krieg}}, \ and\
  \bibinfo {author} {\bibfnamefont {P.}~\bibnamefont {Kopietz}},\ }\href
  {\doibase 10.1103/PhysRevA.93.033609} {\bibfield  {journal} {\bibinfo
  {journal} {Phys. Rev. A}\ }\textbf {\bibinfo {volume} {93}},\ \bibinfo
  {pages} {033609} (\bibinfo {year} {2016})}\BibitemShut {NoStop}%
\bibitem [{\citenamefont {Ticknor}(2009)}]{Ticknor2009PRA}%
  \BibitemOpen
  \bibfield  {author} {\bibinfo {author} {\bibfnamefont {C.}~\bibnamefont
  {Ticknor}},\ }\href {\doibase 10.1103/PhysRevA.80.052702} {\bibfield
  {journal} {\bibinfo  {journal} {Phys. Rev. A}\ }\textbf {\bibinfo {volume}
  {80}},\ \bibinfo {pages} {052702} (\bibinfo {year} {2009})}\BibitemShut
  {NoStop}%
\bibitem [{\citenamefont {Schick}(1971)}]{Schick1971PRA}%
  \BibitemOpen
  \bibfield  {author} {\bibinfo {author} {\bibfnamefont {M.}~\bibnamefont
  {Schick}},\ }\href {\doibase 10.1103/PhysRevA.3.1067} {\bibfield  {journal}
  {\bibinfo  {journal} {Phys. Rev. A}\ }\textbf {\bibinfo {volume} {3}},\
  \bibinfo {pages} {1067} (\bibinfo {year} {1971})}\BibitemShut {NoStop}%
\bibitem [{\citenamefont {Astrakharchik}\ \emph {et~al.}(2009)\citenamefont
  {Astrakharchik}, \citenamefont {Boronat}, \citenamefont {Casulleras},
  \citenamefont {Kurbakov},\ and\ \citenamefont
  {Lozovik}}]{Astrakharchik2009PRA}%
  \BibitemOpen
  \bibfield  {author} {\bibinfo {author} {\bibfnamefont {G.~E.}\ \bibnamefont
  {Astrakharchik}}, \bibinfo {author} {\bibfnamefont {J.}~\bibnamefont
  {Boronat}}, \bibinfo {author} {\bibfnamefont {J.}~\bibnamefont {Casulleras}},
  \bibinfo {author} {\bibfnamefont {I.~L.}\ \bibnamefont {Kurbakov}}, \ and\
  \bibinfo {author} {\bibfnamefont {Y.~E.}\ \bibnamefont {Lozovik}},\ }\href
  {\doibase 10.1103/PhysRevA.79.051602} {\bibfield  {journal} {\bibinfo
  {journal} {Phys. Rev. A}\ }\textbf {\bibinfo {volume} {79}},\ \bibinfo
  {pages} {051602} (\bibinfo {year} {2009})}\BibitemShut {NoStop}%
\bibitem [{\citenamefont {Bertaina}(2013)}]{Bertaina2013EPJ}%
  \BibitemOpen
  \bibfield  {author} {\bibinfo {author} {\bibfnamefont {G.}~\bibnamefont
  {Bertaina}},\ }\href {\doibase 10.1140/epjst/e2013-01763-9} {\bibfield
  {journal} {\bibinfo  {journal} {Eur. Phys. J. Spec. Top.}\ }\textbf {\bibinfo
  {volume} {217}},\ \bibinfo {pages} {153} (\bibinfo {year}
  {2013})}\BibitemShut {NoStop}%
\bibitem [{\citenamefont {Pilati}\ \emph {et~al.}(2005)\citenamefont {Pilati},
  \citenamefont {Boronat}, \citenamefont {Casulleras},\ and\ \citenamefont
  {Giorgini}}]{Pilati2005PRA}%
  \BibitemOpen
  \bibfield  {author} {\bibinfo {author} {\bibfnamefont {S.}~\bibnamefont
  {Pilati}}, \bibinfo {author} {\bibfnamefont {J.}~\bibnamefont {Boronat}},
  \bibinfo {author} {\bibfnamefont {J.}~\bibnamefont {Casulleras}}, \ and\
  \bibinfo {author} {\bibfnamefont {S.}~\bibnamefont {Giorgini}},\ }\href
  {\doibase 10.1103/PhysRevA.71.023605} {\bibfield  {journal} {\bibinfo
  {journal} {Phys. Rev. A}\ }\textbf {\bibinfo {volume} {71}},\ \bibinfo
  {pages} {023605} (\bibinfo {year} {2005})}\BibitemShut {NoStop}%
\bibitem [{\citenamefont {Astrakharchik}\ \emph
  {et~al.}(2007{\natexlab{b}})\citenamefont {Astrakharchik}, \citenamefont
  {Boronat}, \citenamefont {Casulleras}, \citenamefont {Kurbakov},\ and\
  \citenamefont {Lozovik}}]{Astrakharchik2007PRA}%
  \BibitemOpen
  \bibfield  {author} {\bibinfo {author} {\bibfnamefont {G.~E.}\ \bibnamefont
  {Astrakharchik}}, \bibinfo {author} {\bibfnamefont {J.}~\bibnamefont
  {Boronat}}, \bibinfo {author} {\bibfnamefont {J.}~\bibnamefont {Casulleras}},
  \bibinfo {author} {\bibfnamefont {I.~L.}\ \bibnamefont {Kurbakov}}, \ and\
  \bibinfo {author} {\bibfnamefont {Y.~E.}\ \bibnamefont {Lozovik}},\ }\href
  {\doibase 10.1103/PhysRevA.75.063630} {\bibfield  {journal} {\bibinfo
  {journal} {Phys. Rev. A}\ }\textbf {\bibinfo {volume} {75}},\ \bibinfo
  {pages} {063630} (\bibinfo {year} {2007}{\natexlab{b}})}\BibitemShut
  {NoStop}%
\bibitem [{\citenamefont {Giuliani}\ and\ \citenamefont
  {Vignale}(2005)}]{Giuliani2005}%
  \BibitemOpen
  \bibfield  {author} {\bibinfo {author} {\bibfnamefont {G.}~\bibnamefont
  {Giuliani}}\ and\ \bibinfo {author} {\bibfnamefont {G.}~\bibnamefont
  {Vignale}},\ }\href {\doibase 10.1017/CBO9780511619915} {\emph {\bibinfo
  {title} {{Quantum Theory of the Electron Liquid}}}}\ (\bibinfo  {publisher}
  {Cambridge University Press},\ \bibinfo {year} {2005})\BibitemShut {NoStop}%
\bibitem [{\citenamefont {Werner}\ and\ \citenamefont
  {Castin}(2012)}]{Werner2012PRA_1}%
  \BibitemOpen
  \bibfield  {author} {\bibinfo {author} {\bibfnamefont {F.}~\bibnamefont
  {Werner}}\ and\ \bibinfo {author} {\bibfnamefont {Y.}~\bibnamefont
  {Castin}},\ }\href {\doibase 10.1103/PhysRevA.86.013626} {\bibfield
  {journal} {\bibinfo  {journal} {Phys. Rev. A}\ }\textbf {\bibinfo {volume}
  {86}},\ \bibinfo {pages} {013626} (\bibinfo {year} {2012})}\BibitemShut
  {NoStop}%
\bibitem [{\citenamefont {Pandharipande}\ and\ \citenamefont
  {Bethe}(1973)}]{Pandharipande1973PRC}%
  \BibitemOpen
  \bibfield  {author} {\bibinfo {author} {\bibfnamefont {V.~R.}\ \bibnamefont
  {Pandharipande}}\ and\ \bibinfo {author} {\bibfnamefont {H.~A.}\ \bibnamefont
  {Bethe}},\ }\href {\doibase 10.1103/PhysRevC.7.1312} {\bibfield  {journal}
  {\bibinfo  {journal} {Phys. Rev. C}\ }\textbf {\bibinfo {volume} {7}},\
  \bibinfo {pages} {1312} (\bibinfo {year} {1973})}\BibitemShut {NoStop}%
\bibitem [{\citenamefont {Boronat}\ and\ \citenamefont
  {Casulleras}(1999)}]{Boronat1999PRB}%
  \BibitemOpen
  \bibfield  {author} {\bibinfo {author} {\bibfnamefont {J.}~\bibnamefont
  {Boronat}}\ and\ \bibinfo {author} {\bibfnamefont {J.}~\bibnamefont
  {Casulleras}},\ }\href {\doibase 10.1103/PhysRevB.59.8844} {\bibfield
  {journal} {\bibinfo  {journal} {Phys. Rev. B}\ }\textbf {\bibinfo {volume}
  {59}},\ \bibinfo {pages} {8844} (\bibinfo {year} {1999})}\BibitemShut
  {NoStop}%
\bibitem [{\citenamefont {Mora}\ and\ \citenamefont
  {Waintal}(2007)}]{Mora2007PRL}%
  \BibitemOpen
  \bibfield  {author} {\bibinfo {author} {\bibfnamefont {C.}~\bibnamefont
  {Mora}}\ and\ \bibinfo {author} {\bibfnamefont {X.}~\bibnamefont {Waintal}},\
  }\href {\doibase 10.1103/PhysRevLett.99.030403} {\bibfield  {journal}
  {\bibinfo  {journal} {Phys. Rev. Lett.}\ }\textbf {\bibinfo {volume} {99}},\
  \bibinfo {pages} {030403} (\bibinfo {year} {2007})}\BibitemShut {NoStop}%
\bibitem [{\citenamefont {Temple}\ and\ \citenamefont
  {Chapman}(1928)}]{Temple1928PRSL}%
  \BibitemOpen
  \bibfield  {author} {\bibinfo {author} {\bibfnamefont {G.}~\bibnamefont
  {Temple}}\ and\ \bibinfo {author} {\bibfnamefont {S.}~\bibnamefont
  {Chapman}},\ }\href {\doibase 10.1098/rspa.1928.0098} {\bibfield  {journal}
  {\bibinfo  {journal} {Proc. R. Soc. London, Ser. A}\ }\textbf {\bibinfo
  {volume} {119}},\ \bibinfo {pages} {276} (\bibinfo {year}
  {1928})}\BibitemShut {NoStop}%
\bibitem [{\citenamefont {Weinstein}(1934)}]{Weinstein1934PNAS}%
  \BibitemOpen
  \bibfield  {author} {\bibinfo {author} {\bibfnamefont {D.~H.}\ \bibnamefont
  {Weinstein}},\ }\href {\doibase 10.1073/pnas.20.9.529} {\bibfield  {journal}
  {\bibinfo  {journal} {Proc. Nat. Acad. Sci.}\ }\textbf {\bibinfo {volume}
  {20}},\ \bibinfo {pages} {529} (\bibinfo {year} {1934})}\BibitemShut
  {NoStop}%
\bibitem [{\citenamefont {Goedecker}\ and\ \citenamefont
  {Maschke}(1991)}]{Goedecker1991PRB}%
  \BibitemOpen
  \bibfield  {author} {\bibinfo {author} {\bibfnamefont {S.}~\bibnamefont
  {Goedecker}}\ and\ \bibinfo {author} {\bibfnamefont {K.}~\bibnamefont
  {Maschke}},\ }\href {\doibase 10.1103/PhysRevB.44.10365} {\bibfield
  {journal} {\bibinfo  {journal} {Phys. Rev. B}\ }\textbf {\bibinfo {volume}
  {44}},\ \bibinfo {pages} {10365} (\bibinfo {year} {1991})}\BibitemShut
  {NoStop}%
\bibitem [{\citenamefont {Comparin}\ \emph {et~al.}(2018)\citenamefont
  {Comparin}, \citenamefont {Bombin}, \citenamefont {Holzmann}, \citenamefont
  {Mazzanti}, \citenamefont {Boronat},\ and\ \citenamefont
  {Giorgini}}]{Comparin2018Zenodo}%
  \BibitemOpen
  \bibfield  {author} {\bibinfo {author} {\bibfnamefont {T.}~\bibnamefont
  {Comparin}}, \bibinfo {author} {\bibfnamefont {R.}~\bibnamefont {Bombin}},
  \bibinfo {author} {\bibfnamefont {M.}~\bibnamefont {Holzmann}}, \bibinfo
  {author} {\bibfnamefont {F.}~\bibnamefont {Mazzanti}}, \bibinfo {author}
  {\bibfnamefont {J.}~\bibnamefont {Boronat}}, \ and\ \bibinfo {author}
  {\bibfnamefont {S.}~\bibnamefont {Giorgini}},\ }\href {\doibase
  10.5281/zenodo.2425856} {\enquote {\bibinfo {title} {{Data 2D Fermi
  dipoles}},}\ } (\bibinfo {year} {2018}),\ \bibinfo {note}
  {{Zenodo}}\BibitemShut {NoStop}%
\bibitem [{\citenamefont {Park}\ \emph {et~al.}(2015)\citenamefont {Park},
  \citenamefont {Will},\ and\ \citenamefont {Zwierlein}}]{Park2015PRL}%
  \BibitemOpen
  \bibfield  {author} {\bibinfo {author} {\bibfnamefont {J.~W.}\ \bibnamefont
  {Park}}, \bibinfo {author} {\bibfnamefont {S.~A.}\ \bibnamefont {Will}}, \
  and\ \bibinfo {author} {\bibfnamefont {M.~W.}\ \bibnamefont {Zwierlein}},\
  }\href {\doibase 10.1103/PhysRevLett.114.205302} {\bibfield  {journal}
  {\bibinfo  {journal} {Phys. Rev. Lett.}\ }\textbf {\bibinfo {volume} {114}},\
  \bibinfo {pages} {205302} (\bibinfo {year} {2015})}\BibitemShut {NoStop}%
\bibitem [{\citenamefont {Natoli}\ and\ \citenamefont
  {Ceperley}(1995)}]{Natoli1995JCP}%
  \BibitemOpen
  \bibfield  {author} {\bibinfo {author} {\bibfnamefont {V.}~\bibnamefont
  {Natoli}}\ and\ \bibinfo {author} {\bibfnamefont {D.~M.}\ \bibnamefont
  {Ceperley}},\ }\href {\doibase 10.1006/jcph.1995.1054} {\bibfield  {journal}
  {\bibinfo  {journal} {J. Comput. Phys.}\ }\textbf {\bibinfo {volume} {117}},\
  \bibinfo {pages} {171} (\bibinfo {year} {1995})}\BibitemShut {NoStop}%
\bibitem [{\citenamefont {Holzmann}\ and\ \citenamefont
  {Bernu}(2005)}]{Holzmann2005JCP}%
  \BibitemOpen
  \bibfield  {author} {\bibinfo {author} {\bibfnamefont {M.}~\bibnamefont
  {Holzmann}}\ and\ \bibinfo {author} {\bibfnamefont {B.}~\bibnamefont
  {Bernu}},\ }\href {\doibase 10.1016/j.jcp.2004.11.037} {\bibfield  {journal}
  {\bibinfo  {journal} {J. Comput. Phys.}\ }\textbf {\bibinfo {volume} {206}},\
  \bibinfo {pages} {111} (\bibinfo {year} {2005})}\BibitemShut {NoStop}%
\end{thebibliography}%

\end{document}